\documentclass[graybox]{svmult}

\usepackage{mathptmx}       % selects Times Roman as basic font
\usepackage{helvet}         % selects Helvetica as sans-serif font
\usepackage{courier}        % selects Courier as typewriter font
\usepackage{type1cm}        % activate if the above 3 fonts are
\usepackage{makeidx}         % allows index generation
\usepackage{graphicx}        % standard LaTeX graphics tool
\usepackage{multicol}        % used for the two-column index
\usepackage[bottom]{footmisc}% places footnotes at page bottom

\makeindex             % used for the subject index
\begin{document}

\title*{Surface electrons at plasma walls}
\author{Rafael Leslie Heinisch, Franz Xaver Bronold and Holger Fehske}
\institute{Rafael Leslie Heinisch, Franz Xaver Bronold, and Holger Fehske \at Institut f\"ur Physik, Universit\"at Greifswald, 17487 Greifswald, \email{heinisch@physik.uni-greifswald.de, bronold@physik.uni-greifswald.de, and fehske@physik.uni-greifswald.de}}
%\and  \at Institut f\"ur Physik, Universit\"at Greifswald, 17487 Greifswald \email{}
%\and Holger Fehske \at Institut f\"ur Physik, Universit\"at Greifswald, 17487 Greifswald \email{}}
\maketitle

\abstract{In this chapter we introduce a microscopic modelling of the surplus electrons on the plasma wall which complements the classical description of the plasma sheath. First we introduce a model for the electron surface layer to  study the quasistationary electron distribution and the potential at an unbiased plasma wall. Then we calculate sticking coefficients and desorption times for electron trapping in the image states. Finally we study how surplus electrons affect light scattering and how charge signatures offer the possibility of a novel charge measurement for dust grains. }

\section{Introduction}
\label{intro}

When a macroscopic object is brought into contact with an ionised gas its surface is exposed to a fast influx of electrons. Due to the collection of electrons it acquires a negative charge. The wall electrons give rise to a repulsive Coulomb potential which reduces the further influx of electrons until---in quasi-stationarity---electron and ion flux onto the surface balance each other. 
This coincides with the formation of the plasma sheath, an electron depleted region adjacent to the wall.

The traditional modelling of the plasma wall focuses mainly on the plasma sheath which is a macroscopic manifestation of the charge transfer at the plasma boundary. Electrons and ions are usually assumed to annihilate instantly at the wall.
 Thus the wall is considered to be a perfect absorber for the charged species of the plasma. 
This boundary condition implicitly assumes that the charge transfer at the plasma wall occurs on time and length scales too short to be of significance for the development of the discharge. It is thus not necessary to track them \cite{Franklin76}. Improvements to the perfect absorber condition characterise the wall by  an electron-ion  
recombination coefficient and secondary electron emission coefficients for 
various impacting species. 
Still, replacing the electron kinetics by an ad-hoc boundary condition does not allow for a description of the charge transfer at the surface.

\begin{figure}[t]
\includegraphics[width=\linewidth]{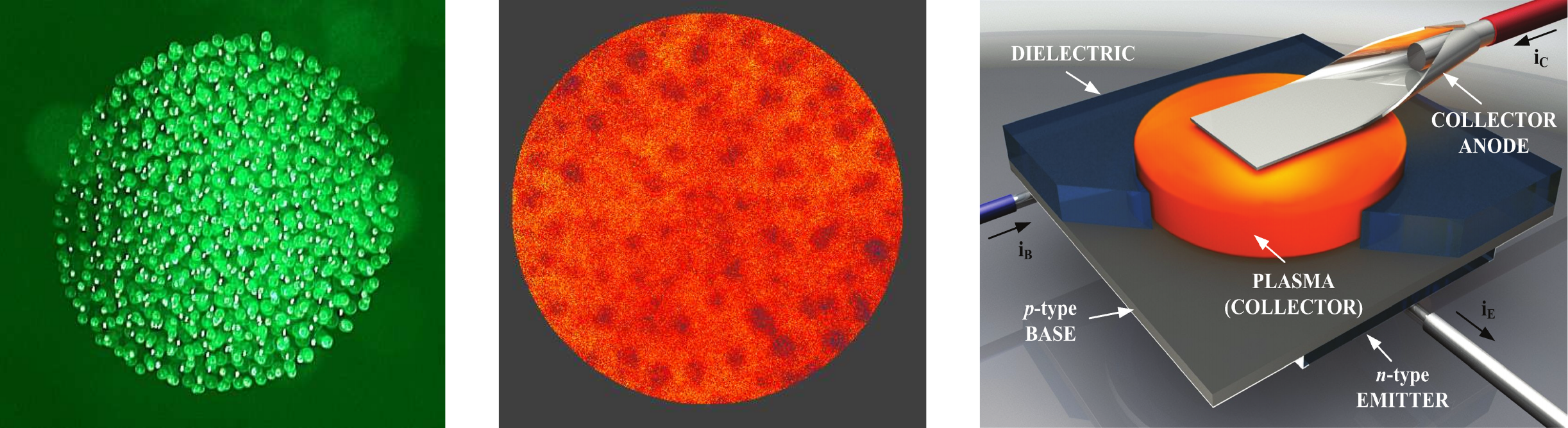}
\caption{Three novel bounded plasmas where surface charges matter. Left: The particle charge controls interactions between dust grains forming a spherical dust ball. Middle: Surface charges (negative charges in blue over neutral background in red) pin the location of filaments for a filamentary dielectric barrier discharge. Right: The plasma serves as a collector in a plasma transistor (reproduced from \cite{WTE10}).}
\label{intro2}
\end{figure}

While this approach may be sufficient for large discharges with small surface to volume ratios it becomes questionable for the modelling of small discharges with large surface to volume ratios. Indeed, the nature and build-up of surface charges have received increasing attention in various novel set-ups of bounded plasmas (illustrated in Fig. \ref{intro2}). 
In dusty plasmas~\cite{BBB09,PAB08,FIK05}, for instance, the charge soaked up by the dust particles controls the overall charge balance of the discharge~\cite{BKS06}. Moreover, the particle charge controls the interaction between dust particles and between dust particles and external electromagnetic fields. The particle charge is thus a central quantity of interest which should be known as precisely as possible.
In dielectric barrier discharges~\cite{GMB02,WYB05,BMG05,SLP07,BWS12,BNW12} or microplasmas \cite{BSE06,Kushner05} surface charges determine the spatio-temporal structure of the discharge~\cite{Kogelschatz03}. Surface charges, for instance, may pin the filaments in the filamentary mode of the dielectric barrier discharge and give the discharge a memory across several cycles.
Lastly, in solid-state based microdischarges~\cite{THW11,DOL10}, the miniaturisation of the discharge increases the surface to volume ratio to such an extent that the (biased) plasma wall becomes  an integral part of the discharge. The charge transfer from the plasma into the solid and vice versa may even be used explicitly, as for instance in the plasma transistor \cite{WTE10}.

\begin{figure}[t]
\includegraphics[width=\linewidth]{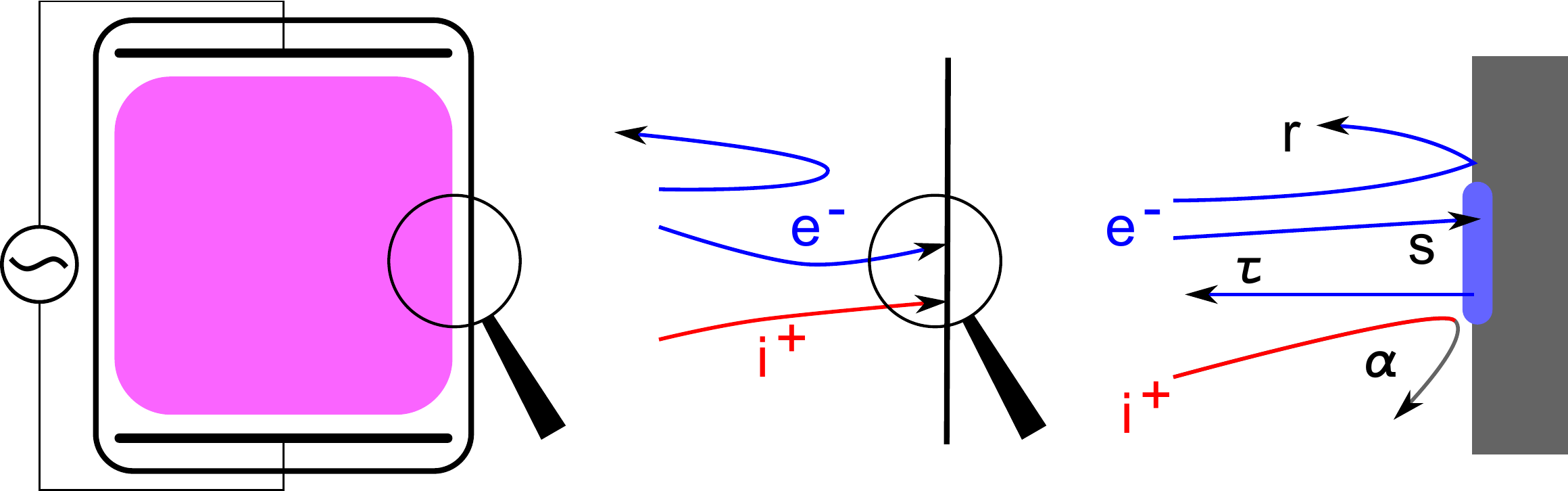}
\caption{Illustration of the plasma and its boundary. The bulk plasma (left) is characterised by quasi-neutrality. In the sheath ions outweigh electrons, quasi-neutrality does not hold and a potential gradient is formed. Note that the bulk as well as the sheath have macroscopic extent and can be described classically. The microscopic processes at the 
surface, for instance electron reflection, sticking or desorption as well as electron ion recombination occur on a length scale smaller than the shortest collision length of the plasma and have to be described by quantum mechanics.}
\label{fig3}
\end{figure}

While the most prominent effect of charge transfer at the plasma wall is the plasma sheath the microscopic processes responsible for it cannot be captured in any sheath model which is solely based on the long range Coulomb potential and classical mechanics. Inspired by Emeleus and Coulter~\cite{EC87} as well as Behnke and coworkers~\cite{BBD97,GMB02} who envisaged a surface plasma of ions and electrons coupled to the bulk plasma by phenomenological rate equations characterised by sticking coefficients, residence times, and recombination coefficients we consider the build-up, distribution and release of surface electrons from a surface physics perspective.  A sketch of the wall bound-processes which cannot be described in a sheath model is given by Fig. \ref{fig3}. The description of the plasma boundary has thus to be augmented by an interface region---the electron surface layer (ESL)---which encompasses not only the Coulomb potential but also short ranged surface potentials. This gives a framework for identifying electron trapping sites, determining the electron distribution and the potential across the interface in quasi-stationarity and calculating electron sticking coefficients. Moreover it serves us to discuss the optical properties of the electron adsorbate on a dust grain.

In the following section we will  introduce our model for the ESL for an unbiased plasma wall. In section \ref{ephys} we will calculate sticking coefficients and desorption times, that are average residence times at the surface, for the case where electron trapping takes place in the image states in front of the surface. Finally in section \ref{mie} we apply our microscopic model to the electrons on a dust particle in order to study charge signatures in light scattering by a spherical dust particle and propose thereby a novel optical charge measurement.

\section{Electron surface layer}
\label{esl}

After the initial charge-up is completed the plasma wall carries a negative charge. In this section we extend the modelling of the plasma
boundary to the region inside the solid and calculate the
plasma-induced modifications of the potential and charge
distribution of the surface within a quasi-static model. 
Although knowing the potential and
charge distribution across the interface may not be of particular
importance for present day technological plasmas, it is of
fundamental interest from an interface physics point of view. In
addition, considering the plasma wall as an integral part of the
plasma sheath may become critical when the miniaturisation
of solid-state-based plasma devices continues.

The current modelling of the plasma wall focuses on the plasma sheath and gives the potential as well as the electron and ion densities across that region. Our model for the ESL extends this into the solid \cite{HBF12}. Specifically, we calculate the potential as well as the distribution of wall bound electrons. Moreover, we give the spatial position, width and chemical potential of the surplus electrons. For this we use a graded interface potential~\cite{Stern78} to 
interpolate between the sheath and the wall potential and distribute the surplus electrons making 
up the wall charge in this potential under the assumption that at quasi-stationarity 
they are thermalised with the wall~\cite{TD85}.

\begin{figure}[t]
\includegraphics[width=\linewidth]{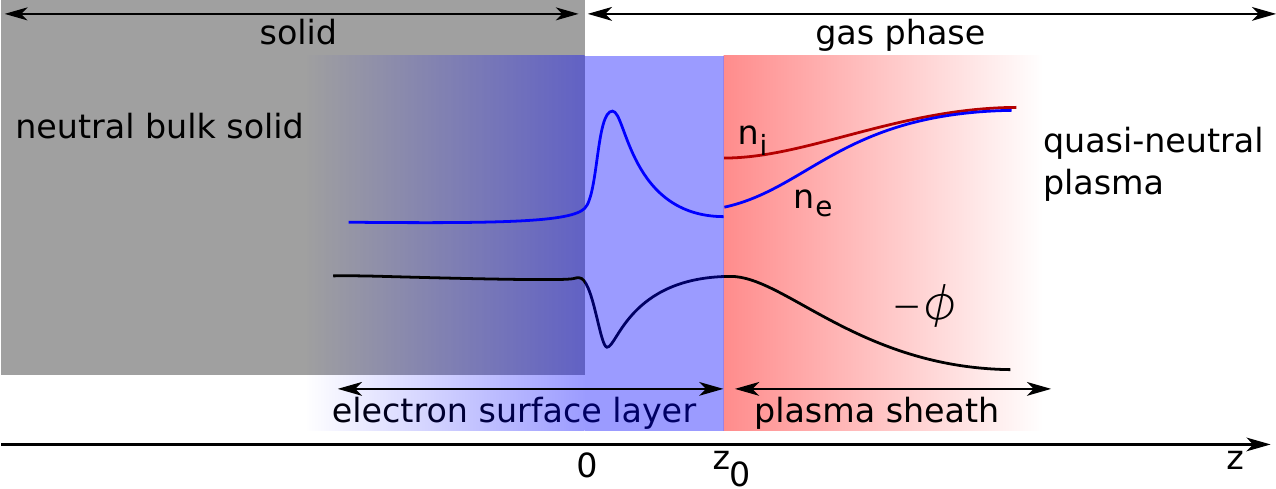}
\caption{Qualitative sketch of the charge double layer formed at a plasma wall. The electrons depleted from the plasma sheath are accumulated in the ESL. Note that the boundary between sheath and ESL is located in front of the crystallographic interface.}
\label{motivationesl}
\end{figure}

As depicted in Fig.~\ref{motivationesl}, we consider an ideal, planar interface at \(z=0\) with the dielectric 
occupying the half-space $z<0$ and the discharge occupying the half-space $z>0$.  At the moment we focus on the 
physical principles controlling the electronic properties of the plasma-wall interface.
Chemical contamination
and structural damage due to the burning gas discharge are discarded. In the model we 
propose the wall charge to be treated as an ESL, which is an interface specific 
electron distribution thermalised with the solid stretching from the plasma sheath 
over the crystallographic interface into the bulk of the dielectric.

The boundary between the ESL and the plasma sheath at \(z_0\) is located in front of the surface. It 
is the position where the attractive force due to the surface potential \(\phi_\mathrm{surf}\) and the 
repulsive force due to the sheath potential \(\phi_\mathrm{sheath}\) balance each other. Thus, \(z_0\) is given by 
\begin{equation}
\phi_\mathrm{surf}^\prime(z_0)+\phi_\mathrm{sheath}^\prime(z_0)=0 . \label{z0condition}
\end{equation}
For \(z<z_0\) an electron is attracted to the surface and thus contained in the ESL while for \(z>z_0\) 
it is repelled back into the plasma. The position $z_0$ is an effective wall for plasma electrons and ions at which the flux balance between electron flux $j_e$ and ion flux$j_i$ has to be fulfilled. In the following we use for simplicity the perfect absorber model \(j_e=j_i\). 
The effective wall position also gives the distance to the surface at which the description of the plasma sheath based on the long-range potential breaks down as the short-range surface potential becomes dominant. On the solid side, for \(z<0\), the ESL is bounded because 
of the shallow potential well formed by the restoring force from the positive charge in the plasma sheath.

Within this quasi-static model the electrons missing in the sheath are accumulated on the surface. For the construction
of our one-dimensional interface model we need the total number per unit area of missing sheath electrons (that 
is, the total surface density of missing sheath electrons) as a function of the wall potential because it 
is this number of electrons which can be distributed across the ESL. Hence, we require a model for the 
plasma sheath.

For simplicity, we use a collision-less sheath model~\cite{Franklin76}, more realistic sheath 
models~\cite{Franklin76,LL05,Riemann91} make no difference in principle. 
In the collision-less sheath electrons are thermalised, 
that is, the electron density \(n_e=n_0 \exp(e \phi/k_B T_e) \), with \(\phi\) the potential, 
\(n_0\) the plasma density and \(T_e\) the electron temperature. The ions enter the sheath 
with a directed velocity \(v_{i0}\) and satisfy a source-free continuity equation, 
\(\mathrm{d}(n_i v_i)/\mathrm{d}z=0\), implying \(n_i v_i =n_0 v_{i0}\), and an equation of 
motion \(M v_i \mathrm{d}v_i/\mathrm{d}z=-e \mathrm{d} \phi /\mathrm{d}z\), with 
\(n_i\) the ion density, and \(M\) the ion mass. The potential \(\phi\) satisfies Poisson's equation 
\(\mathrm{d}^2\phi/\mathrm{d}z^2 =-4\pi e (n_i-n_e)\). Thus, the governing equations for the 
collision-less plasma sheath are~\cite{Franklin76}
\begin{equation}
v_i \frac{\mathrm{d}v_i}{\mathrm{d}z}=-\frac{e}{M} \frac{\mathrm{d} \phi}{\mathrm{d}z} \label{sheath1}
\end{equation}
and
\begin{equation}
\frac{\mathrm{d}^2}{\mathrm{d}z^2}\phi =-4\pi e n_0 \left[ \frac{v_0}{v_i}-\exp \left( \frac{e \phi}{k_B T_e} \right) \right] . \label{sheath2}
\end{equation}
Introducing dimensionless variables \(\eta=-e\phi/(k_BT_e)\), \( \xi=z/\lambda_D\) and \( u=v_i/c_s\),
where \( \lambda_D=\sqrt{k_B T_e/4 \pi n_0 e^2}\) and \(  c_s=\sqrt{k_BT_e/M} \),
equations (\ref{sheath1}) and (\ref{sheath2}) become
\begin{equation}
u u^\prime = \eta^\prime \quad   \mathrm{and} \quad
\eta^{\prime \prime}= \frac{u_0}{u} - \exp(-\eta).
\end{equation}

In the ESL model the plasma occupies not the whole half space \(z>0\) but only the portion
\(z>z_0\) (see Fig.\,\ref{motivationesl}). Integrating the first equation
gives \mbox{\(u=-\sqrt{2\eta+u_0^2}\)}, where \(u_0=v_{i0}/c_s\) is the reduced velocity 
of ions entering the sheath, so that the second equation becomes
\begin{equation}
\eta^{\prime \prime}=-\frac{u_0}{\sqrt{2 \eta +u_0^2}}-\exp(-\eta) . \label{clsheatheq}
\end{equation}
Using the boundary condition that the potential and the field vanish far inside the plasma, that is, 
\(\eta \rightarrow 0\) and \(\eta^\prime \rightarrow 0\) for \(\xi\rightarrow\infty\), Eq. (\ref{clsheatheq}) 
can be integrated once and we obtain
\begin{equation}
\eta^\prime=-\sqrt{-2 u_0 \sqrt{2 \eta +u_0^2}+2 \exp (-\eta) +2 u_0 \sqrt{u_0^2}-2} .
%\label{etaprime}
\end{equation}
For ions entering the sheath with the Bohm velocity \(u_0=-1\). The field at the effective wall 
as a function of the wall potential \(\eta_w=\eta(\xi_0)\) is then given by 
\begin{equation}
\eta_w^\prime=-\sqrt{2\sqrt{2\eta_w+1}+2\exp(-\eta_w)-4} .  \label{etawprime}
\end{equation}

The total surface density of ions in the sheath equals the total surface density of electrons in the ESL \(N\). It can be related by Poisson's equation to the field at the wall and reads 
\begin{equation}
N=\int_{z_0}^\infty \mathrm{d}z (n_i-n_e) =- \frac{1}{4 \pi e} \int_{z_0}^\infty \mathrm{d}z \frac{\mathrm{d}^2\phi}{\mathrm{d}z^2}  =\frac{1}{4\pi e} \frac{\mathrm{d}\phi}{\mathrm{d}z}(z_0)=-n_0 \lambda_D \eta_w^\prime \label{Neq} .
\end{equation}
Combing Eqs.~(\ref{Neq}) and (\ref{etawprime}) gives the total surface density of electrons to be inserted into 
the ESL as a function of the wall potential.

The wall potential itself is determined by the flux balance condition, $j_e=j_i$, which, in the ESL model, 
is assumed to be fulfilled at \(z=z_0\). Using the Bohm flux for the ions and the thermal flux for the electrons,
\begin{equation}
j_i=n_0 \sqrt{\frac{k_B T_e}{M}},\quad   \quad j_e=\frac{1}{4} n_0 \sqrt{\frac{8 k_B T}{\pi m_e}} \exp \left(\frac{e\phi}{k_B T_e}\right) ,
\end{equation}
the wall potential is given by~\cite{Franklin76}
\begin{equation}
\eta_w=\frac{1}{2} \ln \left( \frac{M}{2\pi m_e} \right) ,
\end{equation}
that is,
\begin{equation}
 \phi_w=-\frac{k_B T_e}{2 e} \ln\left(\frac{M}{2 \pi m_e}\right)  .
\end{equation}
In the collision-less sheath model the wall potential depends only on the electron temperature and 
the ion to electron mass ratio. 

\begin{figure}[t]
\includegraphics[width=\linewidth]{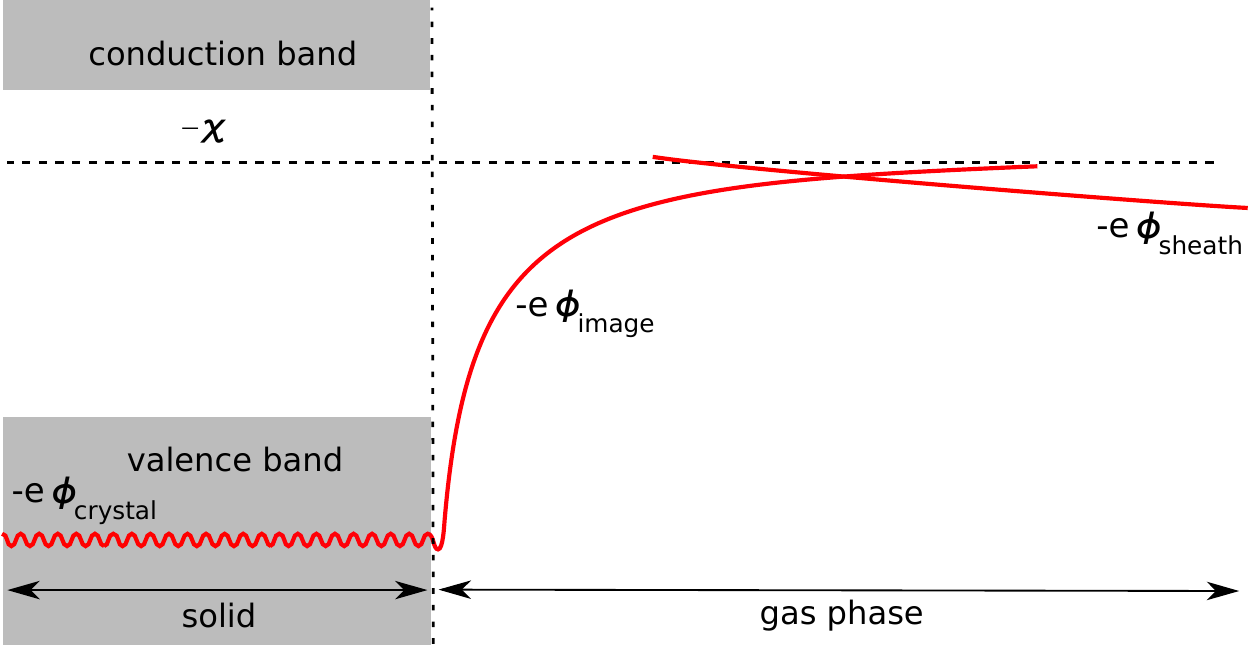}
\caption{Microscopic potentials at the interface between a plasma and a dielectric wall. Shown is  the  microscopic crystal potential merging with the image potential and the sheath potential. Inside the dielectric the crystal potential gives rise to a band structure. }
\label{surfpota}
\end{figure}

\begin{figure}[t]
\includegraphics[width=\linewidth]{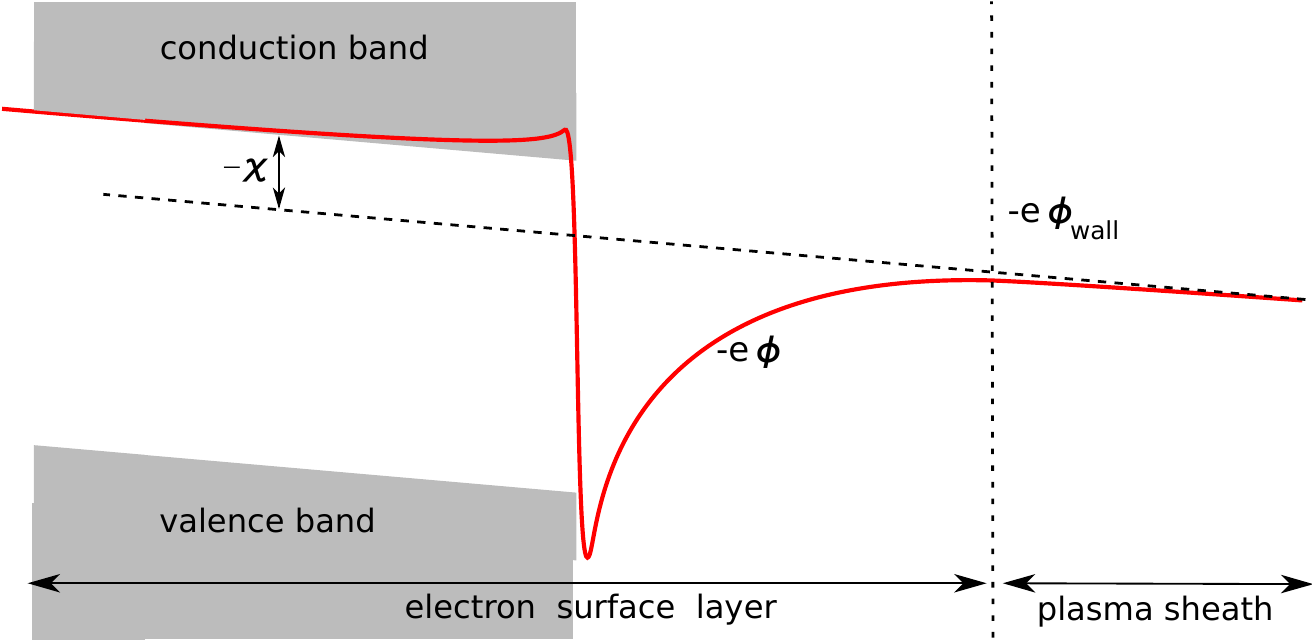}
\caption{Effective potential for the graded interface on which the model of an ESL is based. It encompasses the image potential as well as the potential offset at the interface.}
\label{surfpotb}
\end{figure}

We can now move on to the distribution of surplus electrons in the surface layer. It is primarily determined by the potential in the interface region surrounding an ideal dielectric interface. We first single out the relevant microscopic potentials at the interface (illustrated in Fig. \ref{surfpota}) and then describe an effective model for them (Fig. \ref{surfpotb}).

Far away from the surface the potential is given by the repulsive sheath potential. Closer to the surface an attractive surface potential, the image potential, which stems from the dielectric mismatch at the interface, becomes dominant. 
Classically the image potential is given by the expression \cite{Jackson98} 
\begin{equation}
\phi_\mathrm{im}(z)=\frac{\epsilon_s-1}{4(\epsilon_s+1)}\frac{e}{z} 
\label{ImageClassical}
\end{equation}
where $\epsilon_s$ is the static dielectric constant. 
This expression is valid far away from the wall. Closer to the wall the $1/z$ shape is modified as the potential merges continuously with the crystal potential in the bulk of the dielectric. 
Even more important inside the dielectric is the band structure. For a dielectric the valence band is full while the conduction band is completely empty. Thus surplus electrons can populate the conduction band and the conduction band minimum acts as a long range potential for the electrons. 
In addition to the image potential, the surface potential comprises an offset due to a charge double layer formed immediately at the surface. It is due to electron density leaking out of the solid as a consequence of free energy minimisation or to relaxation and polarisation of the atomic bonds at the surface as a consequence of the truncation of the lattice. This offset of the conduction band minimum at the surface to the potential in  front of it is characterised by the electron affinity of the solid \(\chi\). 
For \(\chi<0\) the conduction band minimum lies above the potential just in front of the solid, for \(\chi>0\) it lies below it.
So far we have considered an uncharged surface. This raises the question whether the electronic structure of the surface is changed by the surplus electrons from the plasma.
 In comparison to the electrons responsible for the chemical binding within
the dielectric the additional electrons coming from the plasma are only a few. The available electronic states and the offset of the energy bands in the 
bulk with respect to the potential outside the dielectric will thus not be changed significantly by the 
presence of the wall charge. 
Note however that a chemical modification of the plasma wall can change the electronic structure. In particular the electron affinity is susceptible to surface coating and adatoms. Surface termination by elements with small electronegativity, for instance hydrogen, may induce a negative electron affinity \cite{CRL98} while elements with larger electronegativity, for instance oxygen, can lead to a positive electron affinity \cite{MRL01}. 

From these microscopic potentials we construct an effective surface potential.
To obtain a realistic image and offset potential without performing an
atomistically accurate calculation we employ the model of a
graded interface.
This model is parametrised by experimentally measured (or where not available theoretically calculated) values for the electron affinity, the dielectric constant, and the conduction band effective mass. First proposed by Stern \cite{Stern78} to remove the unphysical singularity of the image potential at \(z=0\) and later extended to potential offsets at semiconductor heterojunctions \cite{SS84} it assumes the smooth variation of parameters that change abruptly at the interface over a length on the order of the lattice constant.
For our purposes we assume a sinusoidal interpolation of the dielectric constant, offset of the long range potential and effective electron mass with a grading half-width of $d=5\r{A}$ \cite{HBF12}. The model does not account for effects associated with intrinsic surface states (Shockley or Tamm states \cite{Lueth}) or additional states which may arise from the short-range surface potential. Nevertheless the graded interface model is a reasonable description of the surface, suited for dielectrics with ionic bonds which typically have no intrinsic surface states.

The total surface potential 
\begin{equation}
\phi_\mathrm{surf}(z)=\phi_\mathrm{im}(z)+\phi_\mathrm{offset}(z)   \label{empty_surf_pot}
\end{equation} 
comprises the graded image and offset potential.
It is continuous across the crystallographic interface at \(z=0\) and enables us thereby 
to also calculate a smoothly varying electron distribution in the ESL. 

Using Eq.~(\ref{z0condition}) we can now determine the position \(z_0\) of the effective wall, that is, 
the maximum extent of the ESL on the plasma side. The derivative of the  surface potential is 
\(\phi_\mathrm{surf}^\prime =\phi_{\mathrm{ offset}}^\prime+\phi_\mathrm{im}^\prime\).  Due to the
relatively weak field in the sheath compared to the image force, the boundary \(z_0\) will be so far away 
from the interface that \( \phi_{\mathrm{ offset}}^\prime\) vanishes and the image potential 
obeys~(\ref{ImageClassical}). Thus, the boundary between the ESL and the 
plasma sheath is given by 
\begin{equation}
z_0=\sqrt{\frac{(\epsilon_s -1) e}{4 (\epsilon_s +1) \phi_{w}^\prime}} 
\end{equation}
with \(\phi_w^\prime=-(k_BT_e \eta_w^\prime)/(e\lambda_D)\), and \(\eta^\prime_w\) given by~(\ref{etawprime}).

We can now turn to the distribution of the plasma-supplied excess electrons in the ESL. For this quasi-static model we assume that the wall-bound surplus electrons are in thermal equilibrium with the wall. Thus, their distribution minimises the grand canonical potential and satisfies Poisson's equation. 
Inspired by  Tkharev and Danilyuk~\cite{TD85} we apply density 
functional theory~\cite{KS65,Mermin65} to the graded interface. 
For the purpose of this exploratory 
calculation we will use density functional theory in the local approximation.  
Quite generally, the grand canonical potential
of an electron system in an external potential \(V(\vec{r})\) is given in the local approximation by
\begin{equation}
\Omega=\int V(\vec{r}) n(\vec{r}) \mathrm{d}\vec{r} -\frac{e}{2} \int \phi_C(\vec{r}) n(\vec{r}) \mathrm{d}\vec{r}  +G[n]- \mu \int n(\vec{r}) \mathrm{d}\vec{r} ,
\end{equation}
where \(G[n]\) is the grand canonical potential of the homogeneous system with density \(n(\vec{r})\) and the Coulomb potential is determined by
\begin{equation}
\nabla \left( \epsilon(\vec{r}) \nabla \phi_C(\vec{r})\right)=4\pi e n(\vec{r}).
\end{equation}
The ground state electron density minimises \(\Omega\), that is, it satisfies
\begin{equation}
V(\vec{r})-e\phi_C(\vec{r}) + \mu^h(n)-\mu=0~,
\label{DFT}
 \end{equation}
where \(\mu^h(n)=\delta G[n]/\delta n\) is the chemical potential for the homogeneous system. 
 
In our one-dimensional graded interface model Eq.~(\ref{DFT}) reduces to 
\begin{equation}
-e\phi(z)+\mu^h(z)-\mu=0 ,  \label{vareq}
\end{equation}
where \(\mu^h(z)\equiv\mu^h(n(z),T)\), and the electrostatic potential
\begin{equation}
\phi(z)=\phi_\mathrm{surf}(z)+\phi_C(z)  \label{potDFT}
\end{equation}
consists of the potential of the bare surface given by Eq.~(\ref{empty_surf_pot}) and the internal Coulomb potential 
which satisfies Poisson's equation
\begin{equation}
\frac{\mathrm{d}}{\mathrm{d}z}\left( \epsilon(z) \frac{\mathrm{d}}{\mathrm{d}z}\phi_C(z) \right)=4\pi e n(z)   \label{poiseq}
\end{equation}
with the graded dielectric constant \(\epsilon(z)\). The boundary conditions 
\(\phi_C(z_0)=\phi_w\) and \(\phi_C^\prime(z_0)=\phi_w^\prime\)  guarantee continuity of the potential at 
\(z_0\) and  include the restoring force from the positive charge in the sheath. Note that the Coulomb potential derived from this equation already includes the attraction of an electron to the image of the charge distribution. The image potential contains only the self-interaction of an electron to its own image.

For the functional relation \(\mu^h(z)\equiv\mu^h(n(z),T)\) we take the expression adequate for a homogeneous, 
non-interacting, non-degenerate electron gas
\begin{equation}
n(z)=\frac{1}{\sqrt{2}}\left( \frac{m(z) k_B T}{\pi \hbar^2}\right)^\frac{3}{2} e^\frac{\mu^h(z)}{k_BT} \label{neq}
\end{equation}
where \(m(z)\) is the graded effective electron mass.
This is justified because the density of the excess electrons is rather low and the temperature of the surface is 
rather high, typically a few hundred Kelvins. 

For the calculation of the quasi-stationary distribution of surplus electrons Eqs.~(\ref{vareq}) and (\ref{poiseq}) have to be solved iteratively (until \(\mu\) is stationary) subject to the additional constraint \(\int_{z_s}^{z_0} \mathrm{d}z\, n(z)=N\) 
which guarantees charge neutrality between the ESL and the plasma sheath. The position \(z_s<0\) is a cut-off which has to be chosen large enough in order not to affect the numerical 
results. 
In a more refined model for the ESL the crossover of the wall charge to the neutral bulk of the dielectric can be taken into account by splitting the ESL into an interface specific region and a space charge region. This approach  which turns \(z_s\) from an ad-hoc cutt-off into the boundary between the two regions is described in  \cite{HBF12}. For the charge distribution at the interface the simple model is however sufficient.

We now use the ESL model to calculate for a helium discharge in contact with a LiF and Al$_2$O$_3$ surface 
the electron density and potential across the plasma wall. Our main focus lies in the identification
of generic types of electron distributions in the ESL.

\begin{figure}[t]
\includegraphics[width=\linewidth]{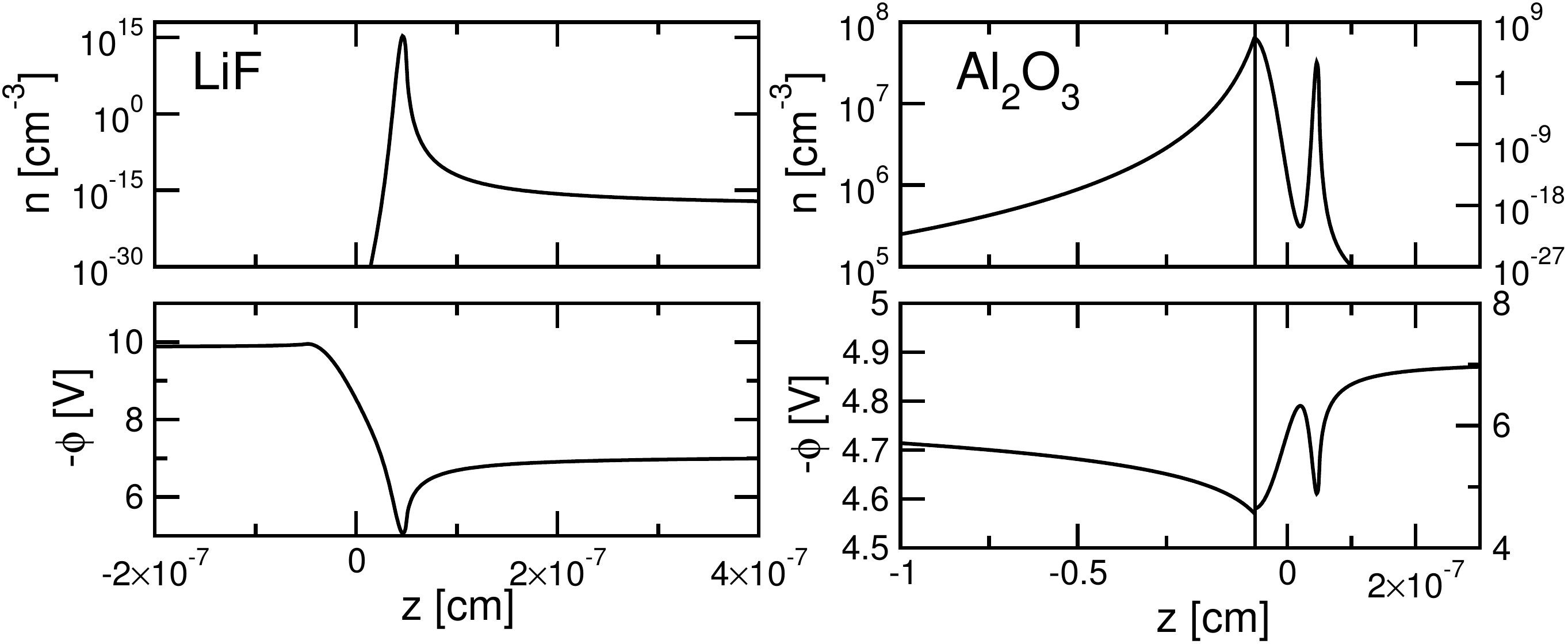}
\caption{Plasma-supplied excess electron density n (upper panel) and the potential $-\phi$ (lower panel) it gives rise to for an LiF (left) and Al$_2$O$_3$ (right) surface in contact with a helium discharge with $n_0=10^7\mathrm{cm}^{-3}$ and $k_BT_e=\mathrm{2eV}$. }
\label{esl_LiF_Al2O3}
\end{figure}

Depending on the electron affinity \(\chi\), one of two scenarios for the distribution of surplus electrons in the ESL may be realised. For negative electron affintiy \(\chi<0\), the conduction band minimum lies above the long-range potential just in front of the surface. It is thus energetically unfavourable for surplus electrons to populate the conduction band. Instead they are located in the image potential in front of the surface. The energy of an electron in the image potential reaches its minimum just in front of the surface at the beginning of the graded interface. This is confirmed by the electron density and the potential for LiF (\(\chi<0\)) shown in the left panel of Fig. \ref{esl_LiF_Al2O3}. 
 For $\chi<0$, the excess electrons coming from the plasma  form therefore an external surface charge in the image potential in 
front of the crystallographic interface. As the electrons do not penetrate into the solid the band bending associated with it is negligible. The external surface charge is very narrow. It can thus be considered as a quasi-two-dimensional electron gas, similar to the surface plasma anticipated by Emeleus and Coulter \cite{EC87,EC88}.
 
For  positive electron affinity \(\chi>0\) the situation is dramatically different. In this case the conduction band minimum is below the potential just outside.
It is thus energetically favourable for electrons to accumulate inside the dielectric. This can be 
seen in the right panel of Fig.~\ref{esl_LiF_Al2O3} which shows the electron density and the potential for an Al$_2$O$_3$ surface (\(\chi > 0\)). 
Although the image potential still induces an attractive well in front of the surface the minimum of the
potential energy for electrons \(-e\phi\) is reached inside the dielectric. Excess electrons coming 
from the plasma are thus mostly located inside the wall where they form a space charge in the conduction band. The electron distribution extends deep into the bulk and entails bending of the energy bands near the surface.
Note the different scales of the axes for the left and right panels of Fig. \ref{esl_LiF_Al2O3}. 
On the scale where variations in the space charge for Al$_2$O$_3$ are noticeable the electron distribution for LiF is basically a vertical line.

So far we have shown the potential and the electron distribution in the ESL. Now, we will compare the potential and the charge 
distribution in the ESL with the ones in the plasma sheath. 
The wall bound electrons in the surface layer constitute  an electron distribution in thermal equilibrium with the wall which balances electron and ion influx at the sheath-ELS boundary $z_0$. On top of this quasistationary electron distribution there are fluxes of electrons and ions, coming from the plasma. These currents, which begin at $z_0$ and persist to the location in the ESL where either electron trapping or electron-ion recombination takes place are not described by our simple ESL model.
 The electron and ion densities in this model are
thus discontinuous at \(z_0\). The potential, however, which has been obtained from the integration of Poisson's equation 
is continuous and differentiable everywhere. Between the crystallographic interface and \(z_0\) the electron and ion 
flux from the plasma would be important. The neglect of the charge densities associated with these fluxes does however 
not affect the potential because they are too small to have a significant effect.

\begin{figure}[t]
\includegraphics[width=\linewidth]{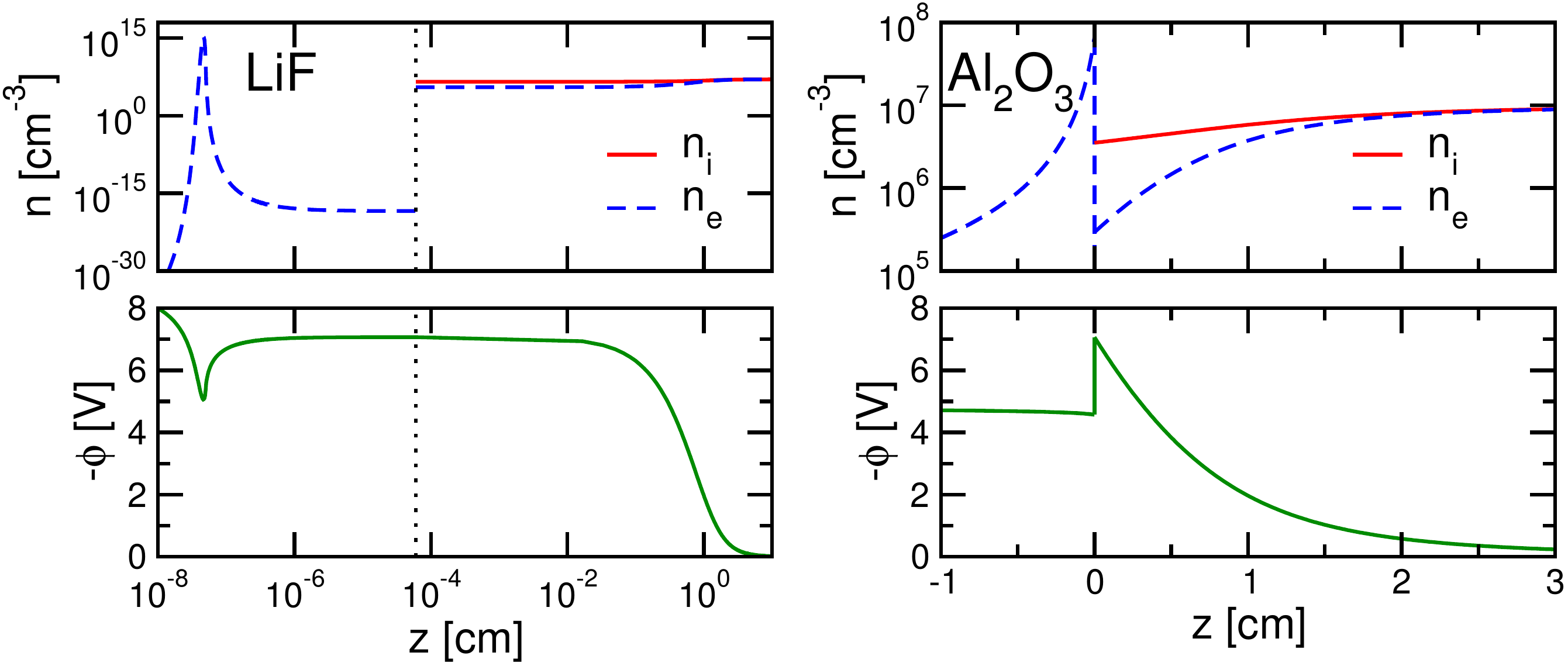}
\caption{Density of plasma-supplied surplus electrons trapped in the ESL, electron and ion density in the plasma sheath, and electric potential for a LiF (left) and an Al$_2$O$_3$ surface (right) in contact with a helium discharge with plasma parameters $n_0=10^7\mathrm{cm}^{-1}$ and $k_BT_e=2eV$ (taken from \cite{BFHM12}). The crystallographic interface is at $z=0$. Note the different scales of the two panels. The deep penetration of the Al$_2$O$_3$ wall charge is due to the neglect of defect states and other collision centres.}
\label{comp_LiF_Al2O3}
\end{figure}

Figure~\ref{comp_LiF_Al2O3} compares the ESL and the plasma sheath for LiF ($\chi<0$, left) and Al$_2$O$_3$ ($\chi>0$, right). We plot the electron and ion densities (upper panel) as well as the electric potential (lower panel). Note the logarithmic scale in the left panel and the linear scale in the right panel. For LiF the surface electrons are bound in the image states in front of the surface due to the negative electron affinity.
Far from the surface, the potential approaches the bulk plasma value chosen to 
be zero. In the sheath the potential develops a Coulomb barrier and reaches the wall potential \(\phi_w\) at \(z_0\), 
the distance where the sheath merges with the surface layer (vertical dotted line). The wall potential is the 
potential just outside to which the energies of the bulk states are 
referenced. Closer to the surface the potential follows the attractive image potential while at the surface the 
repulsive potential due to the negative electron affinity prevents the electron from entering the dielectric
(only scarcely seen on the scale of the figure).

For Al$_2$O$_3$ the excess electrons constituting the wall charge penetrate deep into the dielectric due to the positive electron affinity.
 Compared to the variation of the 
electric potential in the sheath the band bending in the dielectric induced by the wall charge is rather 
small as indicated by the variation of \(\phi\) inside the dielectric. This is because \(\epsilon_s\) is large 
and most surplus electrons are bound close to the crystallographic interface. The interface specific region where the short ranged surface potential varies strongly is very narrow compared to the extent of the two space charge regions.

The ESL should be regarded as the ultimate boundary layer of a bounded gas discharge. Our model for the ESL
captures the coupling of the neutral bulk plasma to an electron plasma at the boundary separated from the bulk
plasma by the plasma sheath. Depending on the electron affinity this boundary plasma can be either a quasi-two-dimensional electron plasma or an electron(-hole) plasma inside the wall.

\section{Electron physisorption}
\label{ephys}

\begin{figure}[t]
\includegraphics[width=\linewidth]{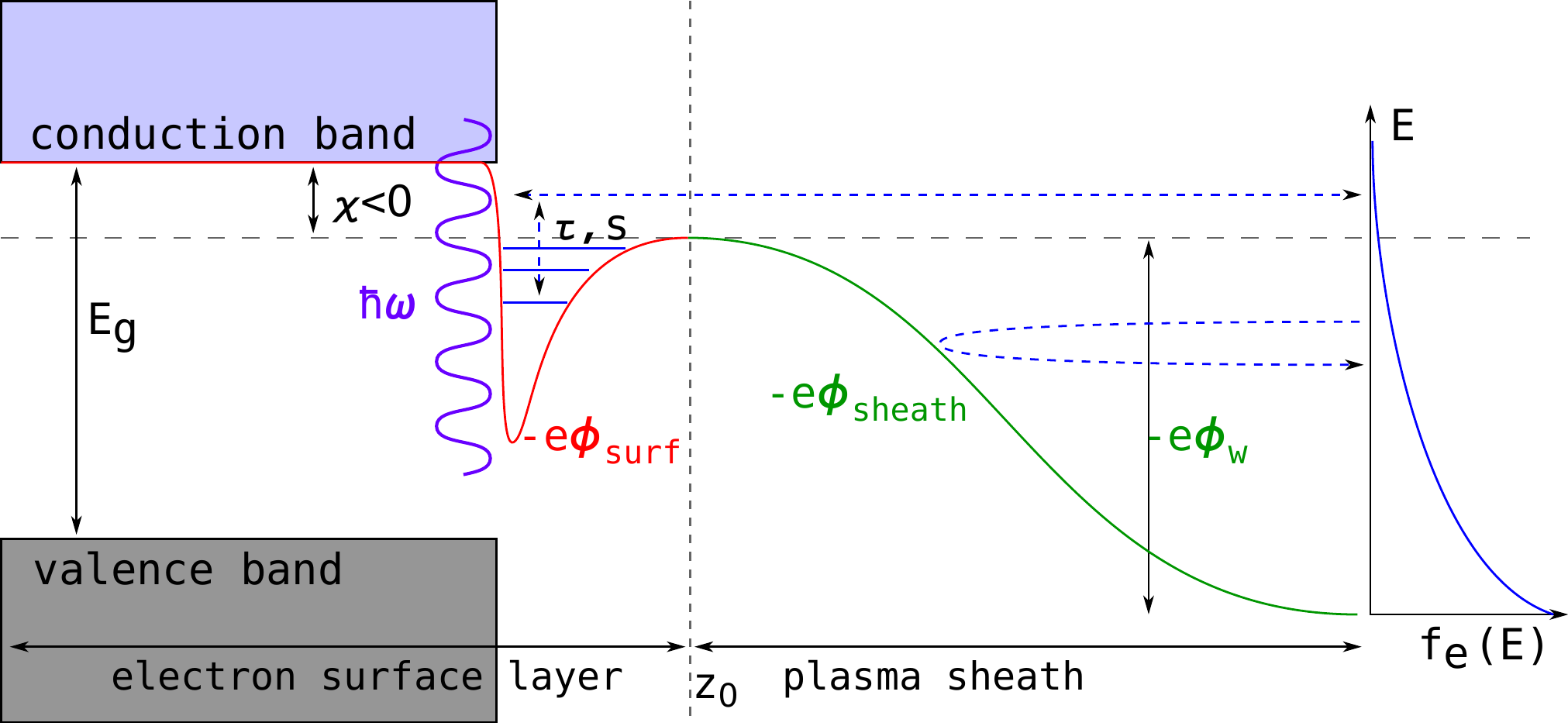}
\caption{Electron trapping in the ESL in front of a dielectric with $\chi<0$.}
\label{esl_to_phys1}
\end{figure}

\begin{figure}[t]
\includegraphics[width=\linewidth]{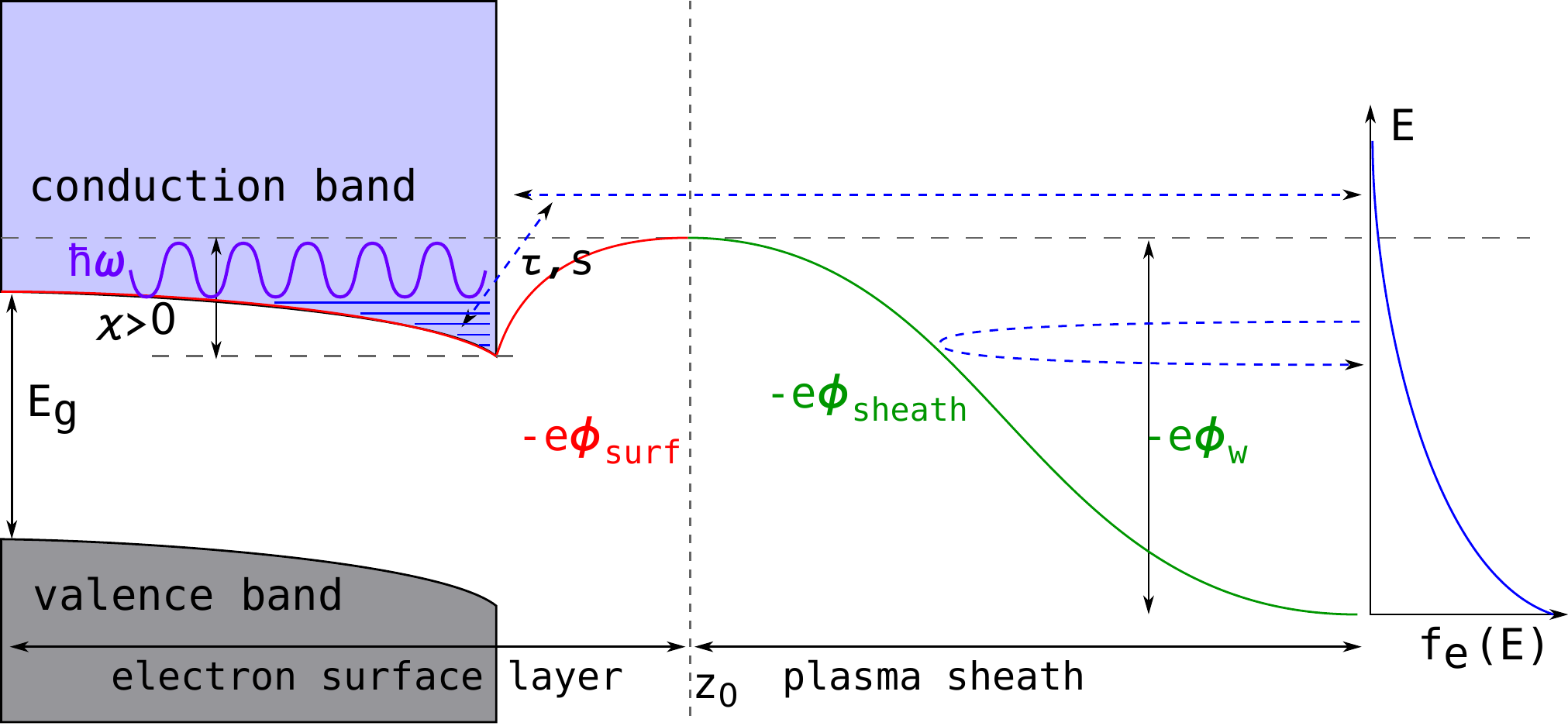}
\caption{Electron trapping in the ESL in a dielectric with $\chi>0$.}
\label{esl_to_phys2}
\end{figure}

We can now turn to the electron kinetics in the ESL. Depending on the electron affinity two distinct scenarios for electron trapping emerge which are schematically shown in Figs. \ref{esl_to_phys1} and \ref{esl_to_phys2}. 
For negative electron affinity (see Fig. \ref{esl_to_phys1}) electrons that overcome the sheath potential do not penetrate the solid as their energy falls in the band gap where no internal states are available. Instead, they are trapped in the image potential in front of the surface. Energy-relaxation is in this case due to surface vibrations which trigger transitions to the bound image states. 
As the image potential is relatively deep compared to the energies of the longitudinal acoustic phonon  responsible for electron trapping and the wavefunctions of the unbound states are suppressed close to the surface the probability for sticking is small. This scenario, which we have investigated  in detail \cite{HBF10a,HBF10b,HBF11}, will be presented below.

For positive electron affinity (see Fig. \ref{esl_to_phys2}) electrons that overcome the sheath potential penetrate into the solid. There they initially occupy high-lying states of the conduction band. Subsequent electron energy relaxation is due to scattering with bulk phonons of the dielectric and the plasma-supplied electrons are eventually trapped at the bottom of the conduction band in the potential well created by the restoring force from the plasma. Due to the high-density of states and an efficient scattering mechanism we expect the sticking coefficient to be much larger in this case. 
However, a detailed analysis of this scenario still remains to be done.

We will now consider the physisorption of an electron in the image potential in front of a dielectric with \(\chi<0\) in detail. 
We describe the time evolution of the occupancy of the 
bound surface states with a quantum-kinetic rate equation \cite{GKT80a,KG86}. It captures all three characteristic 
stages of physisorption: initial trapping, subsequent relaxation and desorption ~\cite{IN76,Brenig82}. 

The time dependence of the occupancies of the bound states is given by \cite{GKT80a,KG86} 
\begin{equation}
\frac{\mathrm{d}}{\mathrm{d}t}n_n(t)=\sum_{n^\prime} \left[W_{n n^\prime} n_{n^\prime}(t) - W_{n^\prime n} n_n(t) \right] 
 -\sum_k W_{k n} n_n(t) +\sum_k \tau_t W_{nk} j_k(t)  \label{fullrateeqn}
 \end{equation}
 which can be rewritten as
 \begin{equation}
\frac{\mathrm{d}}{\mathrm{d}t}n_n(t)= \sum_{n^\prime} T_{nn^\prime} n_{n^\prime}(t)+\sum_k \tau_t W_{nk} j_k(t), \label{fullrateeqn2}
\end{equation}
where \(W_{n^\prime n}\) is the probability per unit time for a transition from a bound state \(n\) to another bound 
state \(n^\prime\), \(W_{kn}\) and \(W_{nk}\) are the probabilities per unit time for a transition from the bound 
state \(n\) to the continuum state \(k\) and vice versa, and \(\tau_t=2L/v_z\) is the transit time through the 
surface potential of width \(L\), which, in the limit \(L\rightarrow \infty\), can be absorbed into the transition 
probability. The matrix \(T_{nn^\prime}\) is defined implicitly by the above equation. The last term in 
Eqs. (\ref{fullrateeqn}) and (\ref{fullrateeqn2}), respectively, gives the increase in the bound state occupancy due 
to the coupling between bound and unbound surface states.

The probability for an approaching electron in the continuum state \(k\) to make a transition to any of the bound states 
is given by the prompt energy-resolved sticking coefficient
\begin{equation}
s_{e,k}^\mathrm{prompt}=\tau_t\sum_n W_{nk} .
\end{equation}

Treating the incident electron flux as an externally specified parameter, the solution of Eq. (\ref{fullrateeqn}) describes the subsequent relaxation and desorption. It is given by
\begin{equation}
n_n(t)=\sum_{\kappa} e^{-\lambda_\kappa t} \int_{-\infty}^t \mathrm{d}t^\prime e^{\lambda_\kappa t^\prime} e_n^{(\kappa)} \sum_{kl} \tilde{e}_l^{(\kappa)} \tau_t W_{lk} j_k(t^\prime)  \label{rateqensolution}
\end{equation}
where \(e_n^{(\kappa)} \) and \(\tilde{e}_n^{(\kappa)}\) are the right and left eigenvectors to the eigenvalue \(-\lambda_\kappa\) of the matrix \(T_{nn^\prime}\).

If the modulus of one eigenvalue, \(\lambda_0\), is considerably smaller than the moduli of the other eigenvalues, 
\(\lambda_\kappa\), a unique desorption time and a unique sticking coefficient can be identified \cite{KG86}. In 
this case \(\lambda_0\) governs the long time behaviour of the equilibrium occupation of the bound states, 
\(n_n^\mathrm{eq} \sim e^{- E_n/k_BT_s}\), and its inverse can be identified with the desorption time, \(\lambda_0^{-1}=\tau_e\). The bound state occupancy  \(n_n(t)\) splits into a slowly varying part \(n_n^0(t)\) given by the \(\kappa =0\) summand in Eq. (\ref{rateqensolution}) and a quickly varying part  \(n_n^f(t)\) given by the sum over \(\kappa \neq0\) in Eq. (\ref{rateqensolution}).

The electron adsorbate, i.e. the fraction of the trapped electron remaining in the surface states for times on the order of the desorption time, is given by the slowly varying part only, \(n^0(t)=\sum_n n_n^0(t)\). 
Differentiating \(n^0(t)\) with respect to the time, 
\begin{equation}
\frac{d}{dt}{n}^0(t)=\sum_k s_ {e,k}^\mathrm{kinetic} j_k(t) - \lambda_0 n^0(t) 
\end{equation}
we can, following Brenig \cite{Brenig82}, identify the kinetic energy-resolved sticking coefficient
\begin{equation}
s_{e,k}^\mathrm{kinetic}=\tau_t \sum_{n,n^\prime}  e_{n^\prime}^{(0)} \tilde{e}_n^{(0)} W_{n k}
\end{equation}
giving the probability for both, initial trapping and subsequent relaxation. 

If the incident unit electron flux corresponds to an electron with Boltzmann distributed kinetic energies, the prompt or kinetic energy-averaged sticking coefficient is given by
\begin{equation}
s_e^{\dots}=\frac{\sum_k s_{e,k}^{\dots} k e^{-\beta_e E_k}}{\sum_k k e^{-\beta_e E_k}},
\end{equation}
where \(\beta_e^{-1}=k_B T_e\) is the mean electron energy.

\begin{figure}[t]
\sidecaption
\includegraphics[width=0.5\linewidth]{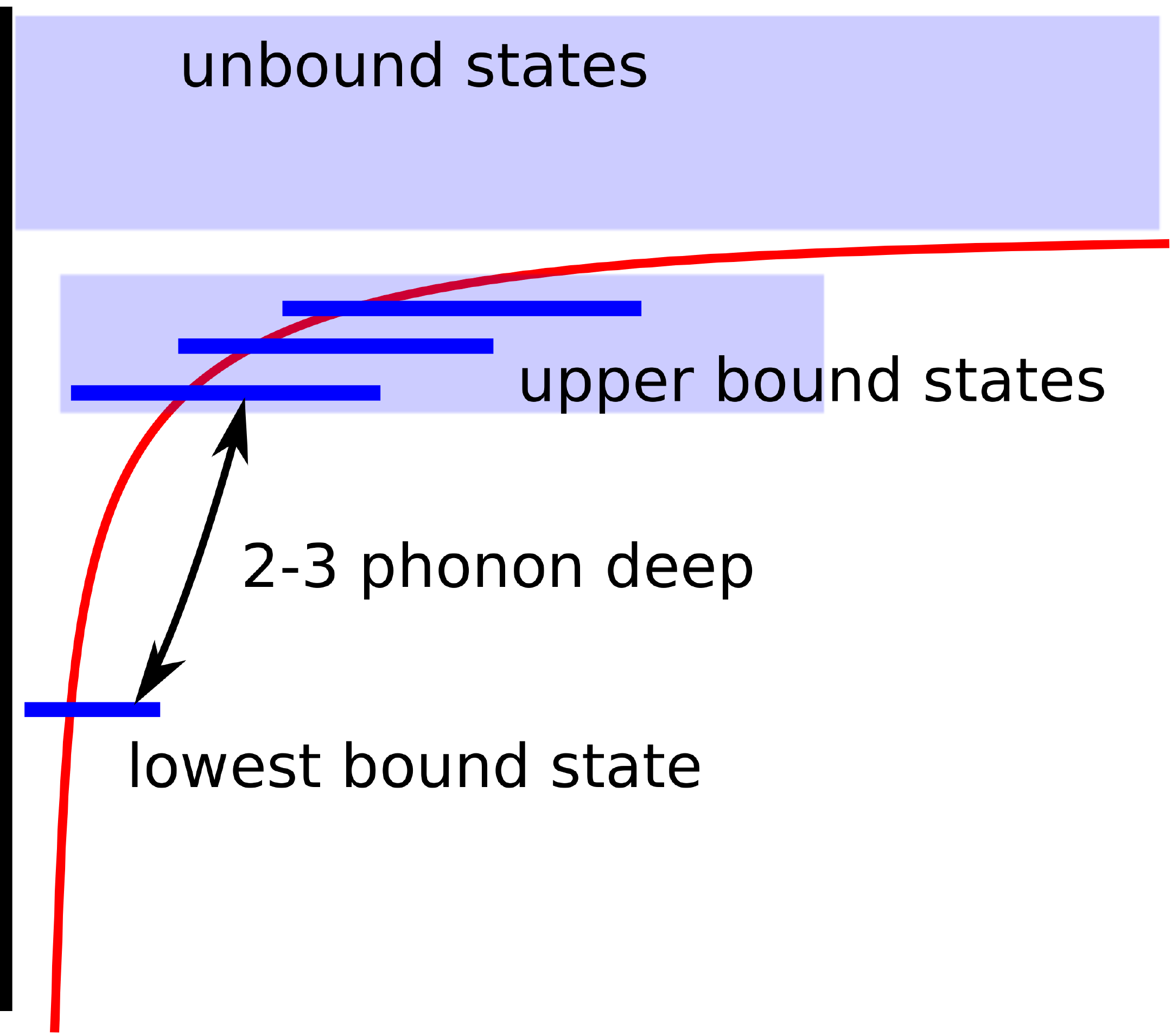}
\caption{Schematic representation of the series of image states in the image potential. The lowest bound states lies very deep and is connected to a group of upper bound states only by a two or three- phonon processes. The upper bound states are connected by one-phonon processes to the continuum.}
\label{ipotential2}
\end{figure}

These kinetic equations rely on the knowledge of the transition probabilities. They have 
to be calculated from a microscopic model for the electron-surface interaction. The image potential supports a series of bound states which can trap electrons temporarily at the surface. Transitions between image states are due to dynamic perturbations of the surface potential. The image potential is very steep near the surface. A particularly strong perturbation arises therefore from the surface vibrations induced by the longitudinal acoustic bulk phonon perpendicular to the surface. 
Describing them with a Debye model, the maximum phonon energy of this mode is the Debye energy \(\hbar \omega_D\). 
Measuring energies in units of the Debye energy, important dimensionless parameters 
characterising the potential depth are
\begin{equation}
\epsilon_n=\frac{E_n}{\hbar \omega_D}  \quad \mathrm{ and } \quad  \Delta_{nn^\prime}=\frac{E_n-E_{n^\prime}}{\hbar \omega_D} ,
\end{equation}
where \(E_n<0\) is the energy of the \(\mathrm{n}^\mathrm{th}\) bound state. 

The depth of the image potential can be classified with respect to the Debye energy. If  \(-n+1> \Delta_{12} >- n\), we call the 
potential $n$-phonon deep. If \(\epsilon_1>-1\) we call it shallow. For a shallow potential the lowest bound state is coupled by a one-phonon transition to the continuum. If the potential is \(n\)-phonon deep an \(n\)-phonon process links the lowest bound state to the second bound states, which is linked by one-phonon transitions either directly or via intermediates to the continuum. For the dielectrics which we consider the potential is two-phonon (MgO) or three-phonon deep (LiF). A schematic representation of the image states is given in  Fig. \ref{ipotential2}.

For the calculation of the transition probabilities used in the rate equation we need a microscopic model for the image potential and the electron-surface vibration interaction. Classically the image potential ensues from the dielectric mismatch at the interface. On a microscopic level it arises from the coupling of the electron to a polarisable surface mode of the solid. For a dielectric this is a surface phonon. For LiF and MgO the low frequency dielectric function is dominated by a transverse optical (TO) phonon with frequency $\omega_{TO}$. It can be approximated by \cite{EM73}
\begin{equation}
\epsilon(\omega)=1+(\epsilon_s-1)\frac{\omega_{TO}^2}{\omega_{TO}^2-\omega^2},
\end{equation}
where $\epsilon_s$ is the static dielectric constant. The bulk TO-phonon gives rise to a surface phonon. Its frequency $\omega_s=\omega_{TO}\sqrt{(1+\epsilon_s)/2}$ is determined by the condition $\epsilon(\omega_s)=-1$. The electron couples to this surface phonon according to \cite{EM73,Barton81} 
\begin{equation}
H=-\frac{\hbar^2}{2m}\Delta + \sum_{\mathbf{Q}_\parallel} \hbar \omega_s a_{\mathbf{Q}_\parallel}^\dagger a_{\mathbf{Q}_\parallel} + \sum_{\mathbf{Q}_\parallel} \left( \phi_{\mathbf{Q}_\parallel}(\mathbf{x}_\parallel)a_{\mathbf{Q}_\parallel} + \phi_{\mathbf{Q}_\parallel}^\ast(\mathbf{x}_\parallel)a_{\mathbf{Q}_\parallel}^\dagger \right) \label{EMHamiltonian}
\end{equation}
with
\begin{equation}
\phi_{\mathbf{Q}_\parallel}(\mathbf{x}_\parallel)=\frac{2 \sqrt{\pi \omega_s \Lambda_0 \hbar}e}{Q_\parallel^{1/2}A^{1/2}} e^{-Q_\parallel |z|} e^{i \mathbf{Q}_\parallel \mathbf{x}_\parallel}
\end{equation}
where $\Lambda_0=(\epsilon_s-1)/(4(\epsilon_s+1))$ and the subscript $\parallel$ denotes vectors parallel to the surface and \(a_{\mathbf{Q}_\parallel}^{(\dagger)}\) are annihilation (creation) operators for surface phonons.

Applying a unitary transformation \cite{Barton81} separates the coupling into a static part which takes the classical form of the image potential \(\sim 1/z\) and a dynamic part of the electron-surface phonon coupling, which encodes recoil effects and encompasses momentum relaxation parallel to the surface.
While the classical image potential allows a simple description of the image states which captures their properties fairly well it is  not sufficient for the calculation of probabilities for surface vibration induced transitions, as they perturb the electronic states most strongly close to the surface where the image potential is steepest. 
Unfortunately, in this region the classical image potential has an unphysical divergence. To remedy this we use a variational procedure to extract the static image potential from \cite{EM73}. Thereby we keep some recoil effects which make the recoil-corrected image potential \(\sim 1 /(z + z_c)\) with a cut-off parameter $z_c=\sqrt{\hbar/2 m \omega_s \pi^2}$ \cite{HBF10a} divergence free.

Transitions between the eigenstates of the recoil-corrected
image potential are due to the longitudinal-acoustic bulk phonon perpendicular to the
surface. 
The Hamiltonian from which we calculate the transition probabilities  is given by
\begin{equation}
H=H_{e}^\mathrm{static}+H_{ph}+H_{e-ph}^\mathrm{dyn} 
\end{equation}
where 
\begin{equation}
H_e^\mathrm{static}=\sum_q E_q c_q^\dagger c_q
\end{equation}
describes the electron in the recoil-corrected image potential, 
\begin{equation}
H_{ph}=\sum_Q \hbar \omega_Q b_Q^\dagger b_Q 
\end{equation}
describes the free dynamics of the bulk longitudinal acoustic phonon responsible for transitions between surface states, and
\begin{equation}
H_{e-ph}^\mathrm{dyn}=\sum_{q,q^\prime} \langle q^\prime | V_p(u,z) |q\rangle c_{q^\prime}^\dagger c_q
\end{equation}
denotes the dynamic coupling of the electron to the bulk phonon. 

The perturbation \(V_p(u,z)\) can be identified as the difference between the displaced surface potential and the static surface potential. It reads, after the transformation \(z \rightarrow z- z_c\), 
\begin{equation}
V_p(u,z)=-\frac{e^2 \Lambda_0}{z+u}+\frac{e^2\Lambda_0}{z} . \label{compactperturb} 
\end{equation}
In general, multi-phonon processes can arise both from the nonlinearity of the electron-phonon coupling 
\(H_{e-ph}^\mathrm{dyn}\) as well as from the successive actions of \(H_{e-ph}^\mathrm{dyn}\) encoded in the 
T matrix equation,
\begin{equation}
T=H_{e-ph}^\mathrm{dyn}+H_{e-ph}^\mathrm{dyn} G_0 T \mathrm{ ,}
\end{equation}
where \(G_0\) is given by 
\begin{equation}
G_0=\left(E-H_e^\mathrm{static}-H_{ph}+i0^+\right)^{-1} .
\end{equation}
The transition probability per unit time from an electronic state \(q\) to an electronic state \(q^\prime\) 
encompassing both types of processes is given by~\cite{BY73}
\begin{equation}
W_{q^\prime q}=\frac{2\pi}{\hbar} \sum_{s,s^\prime} \frac{e^{-\beta_s E_s}}{\sum_{s^{\prime\prime}} e^{-\beta_s E_{s^{\prime \prime}}} } \left|\langle s^\prime,q^\prime | T | s,q\rangle \right|^2 \nonumber    \delta (E_s -E_{s^\prime} +E_q -E_{q^\prime}), \label{generalTR}
\end{equation}
where \(\beta_s = (k_B T_s)^{-1}\),  with \(T_s\) the surface temperature and \(|s\rangle\) and \(|s^\prime \rangle\) the initial and final phonon states. We are  only interested in the transitions between electronic states. It is thus natural to average in Eq. (\ref{generalTR}) over all phonon states. The delta function guarantees energy conservation.

In principle, multiphonon transition rates can be obtained by iterating the T-matrix and evaluating Eq. (\ref{generalTR}). Up to \(\mathcal{O}(u^3)\), for instance, the T-matrix reads
\begin{equation}
T=V_1+V_2+V_3+V_1G_0V_1+V_2G_0V_1+V_1G_0V_2+V_1G_0V_1G_0V_1,
\end{equation}
where the \(V_i\sim u^i\) originate from expanding Eq. (\ref{compactperturb}) in the displacement \(u\). The T-matrix enters as $\langle T\rangle \langle T^\ast \rangle$ into the transition probability. The term $\langle V_1 \rangle \langle V_1^\ast \rangle$ can be identified as the Golden Rule transition probability. Proportional to $u^2$ it is a one-phonon process. Two-phonon processes, proportional to $u^4$, are represented by the terms
\begin{eqnarray}
&\langle V_1 \rangle \langle V_3^\ast \rangle, \langle V_3 \rangle \langle  V_1^\ast\rangle, \langle V_1\rangle \langle V_2^\ast G_0^\ast V_1^\ast \rangle, \langle V_1 G_0 V_2 \rangle \langle V_1^\ast \rangle,  \langle V_1 \rangle \langle V_1^\ast G_0^\ast V_2^\ast \rangle,\nonumber \\
& \langle V_2G_0V_1  \rangle \langle V_1^\ast \rangle, \langle V_1\rangle \langle V_1^\ast G_0^\ast V_1^\ast G_0^\ast V_1^\ast \rangle, \langle V_1 G_0 V_1 G_0 V_1 \rangle \langle V_1^\ast \rangle, 
\end{eqnarray}
and
\begin{equation}
\langle V_2 \rangle \langle V_2^\ast \rangle, \langle V_1G_0V_1 \rangle \langle V_1^\ast G_0^\ast V_1^\ast \rangle, \langle V_2 \rangle \langle V_1^\ast G_0^\ast V_1^\ast \rangle, \langle V_1 G_0 V_1 \rangle \langle V_2^\ast \rangle.
\end{equation}
A complete two-phonon calculation would take all these processes into account as they stand. This is however not always necessary. A closer analysis (see  \cite{HBF10a}) of the first group of terms reveals that they are two-phonon corrections to transitions already enabled by a one-phonon process. We assume that theses corrections are small and evaluate only the second group for transitions where they enable two-phonon transitions which are not merely corrections to one-phonon transitions. The details of the evaluation of the transition probabilities, including a regularisation of divergences by taking a finite phonon lifetime into account,  can be found in  \cite{HBF10a,HBF10b}. 

\begin{figure}[t]
\includegraphics[width=\linewidth]{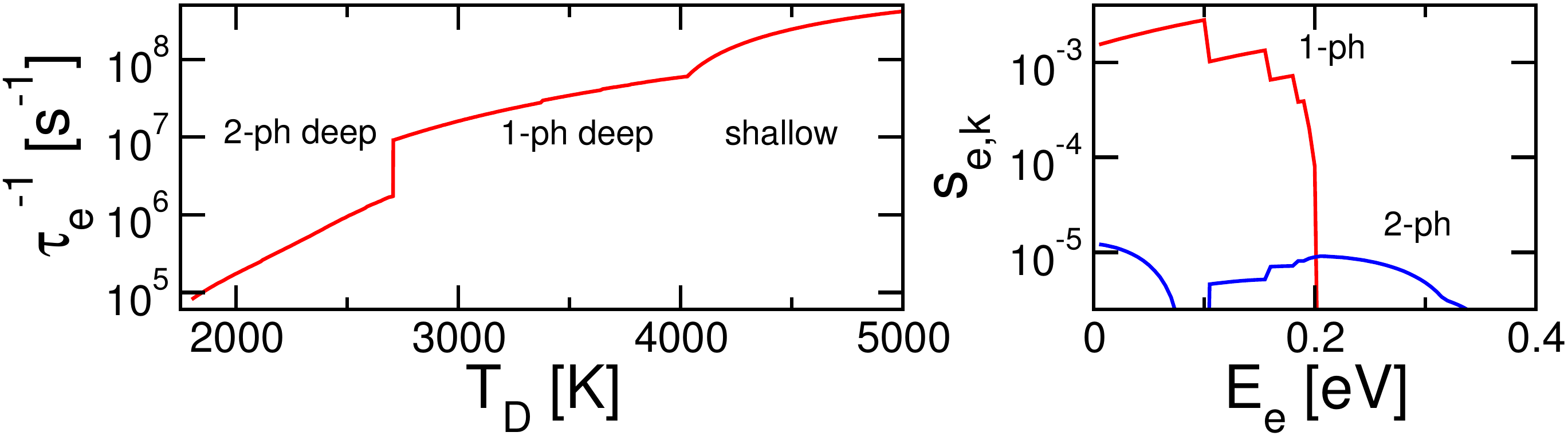}
\caption{Left: Inverse desorption time $\tau_e^{-1}$ in the two-phonon approximation for $T_D/T=5$ as a function of the Debye temperature $T_D$. Data for $T_D=2500K$ apply to graphite. Right: Prompt  energy-resolved sticking coefficient as a function of the electron energy for a two phonon deep potential ($T_D=2500K$ and $T_s=357K$). The contribution of one-phonon processes (red) far outweighs the contribution from two phonon processes (blue). }
\label{stick_des_analysis}
\end{figure}

The expansion of the T-matrix allows the calculation of transition probabilities for the two-phonon deep potentials of graphite and MgO. However,  for a  three-phonon deep potential, for instance for LiF and CaO, this approach is no longer feasible.  From Refs. \cite{HBF10a,HBF10b} we qualitatively know the relevance of different types of multi-phonon processes. For continuum to bound state transitions, for instance, one phonon processes are sufficient at low electron energies. We will therefore compute the transition probabilities between bound and continuum states in the one-phonon approximation. For transitions between bound states, we found that multi-phonon processes due to the nonlinearity of the electron-phonon coupling tend to be more important than the multi-phonon processes due to the iteration of the T-matrix. Hence we expect that an approximation which takes only the nonlinearity of the electron-phonon interaction nonperturbatively into account to be sufficient for the identification of the generic behaviour of multi-phonon-mediated adsorption and desorption. This approach is described in \cite{HBF11}.

We now turn to results for the desorption time and the sticking coefficient for physisorption in the image states. Before showing results for MgO and LiF, we consider the dependence of the electron kinetics on the potential depth and the relevance of one or two-phonon processes. 
For this we show in the left panel of Fig. \ref{stick_des_analysis} the inverse of the desorption time as a function of the Debye temperature \(T_D=\hbar \omega_D/k_B\) which sets the energy scale of the acoustic phonons. While the absolute depth of the potential remains constant, varying the phonon energy tunes the effective potential depth. Fig. \ref{stick_des_analysis} reveals that for a shallow potential desorption is most efficient while for a one-phonon deep potential it gradually becomes less efficient. When the potential becomes two-phonon deep  desorption suddenly becomes even slower which reflects the small magnitude of two-phonon transitions compared to one-phonon transitions. This justifies our approximation of neglecting two-phonon corrections to transitions already enabled by a one-phonon process. For a Debye temperature $T_D=2500K$ the results apply to graphite.

\begin{figure}[t]
\sidecaption
\includegraphics[width=0.6\linewidth]{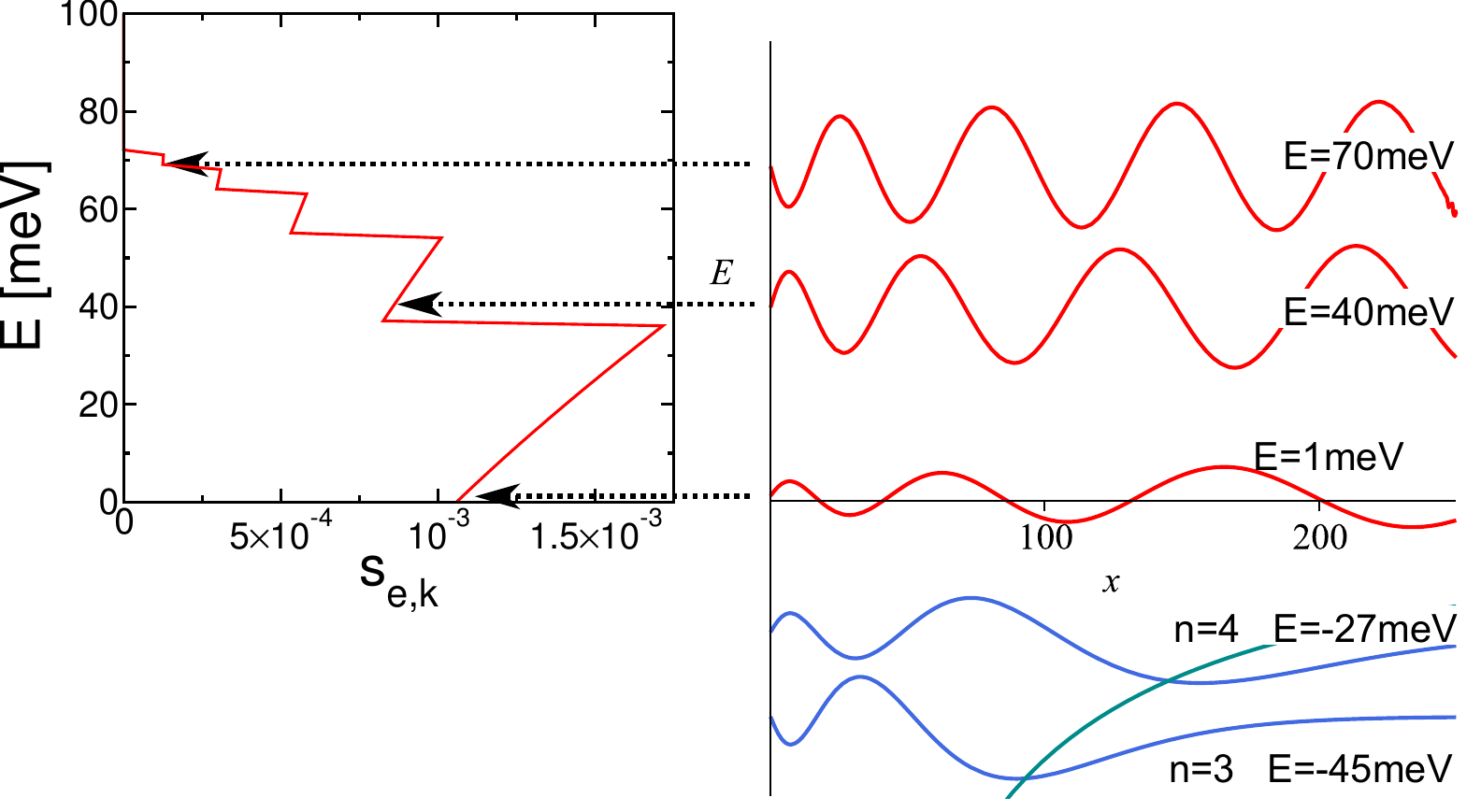}
\caption{Energy resolved prompt sticking coefficient for MgO (left) and schematic representation of the bound and continuum wavefunctions (right). The suppression of the amplitude of the continuum wavefunction close to the surface for small electron energies causes the small sticking coefficient.}
\label{wavefunction_sticking_Mgo}
\end{figure}

The energy-resolved prompt sticking coefficient, plotted in the right panel of Fig. \ref{stick_des_analysis} for graphite, shows that also for bound state-continuum transitions one-phonon processes far outweigh two-phonon processes. Thus, sticking is mainly due to one-phonon processes. Moreover we find that the sticking coefficient drops sharply at specific energies. These accessibility thresholds occur whenever one bound state becomes no longer accessible from the continuum because the energy difference exceeds one Debye energy.

The sticking coefficient takes relatively small values---on the order of $10^{-3}$. This is due to the small matrix elements between bound and continuum states. Fig. \ref{wavefunction_sticking_Mgo} shows the wave functions of representative bound and continuum states for MgO. For large energy the continuum wave functions have sinusoidal shape whereas for small energies their amplitude is significantly suppressed close to the surface. These two behaviours correspond to two limits for the wavefunction. For simplicity we discuss this for a $1/x$ potential without cut-off. The continuum wavefunctions read $\phi_k(x)\sim \frac{1}{\sqrt{k}}M_{\frac{-i}{k},\frac{1}{2}}\left(2ik\Lambda_0 x \right)$ where $x=z/a_B$. For $x\rightarrow \infty$ we obtain $\phi_k(x)\sim \sin (\Lambda_0 k x)$ which also holds for large $k$ while for $k\rightarrow 0$ we obtain $\phi_k(x)\sim \sqrt{k} \sqrt{2 \Lambda_0 x} J_1(\sqrt{8\Lambda_0 x}) $. The proportionality to $\sqrt{k}$ entails a strong suppression of the wavefunction for low energy. 

\begin{figure}[t]
\includegraphics[width=\linewidth]{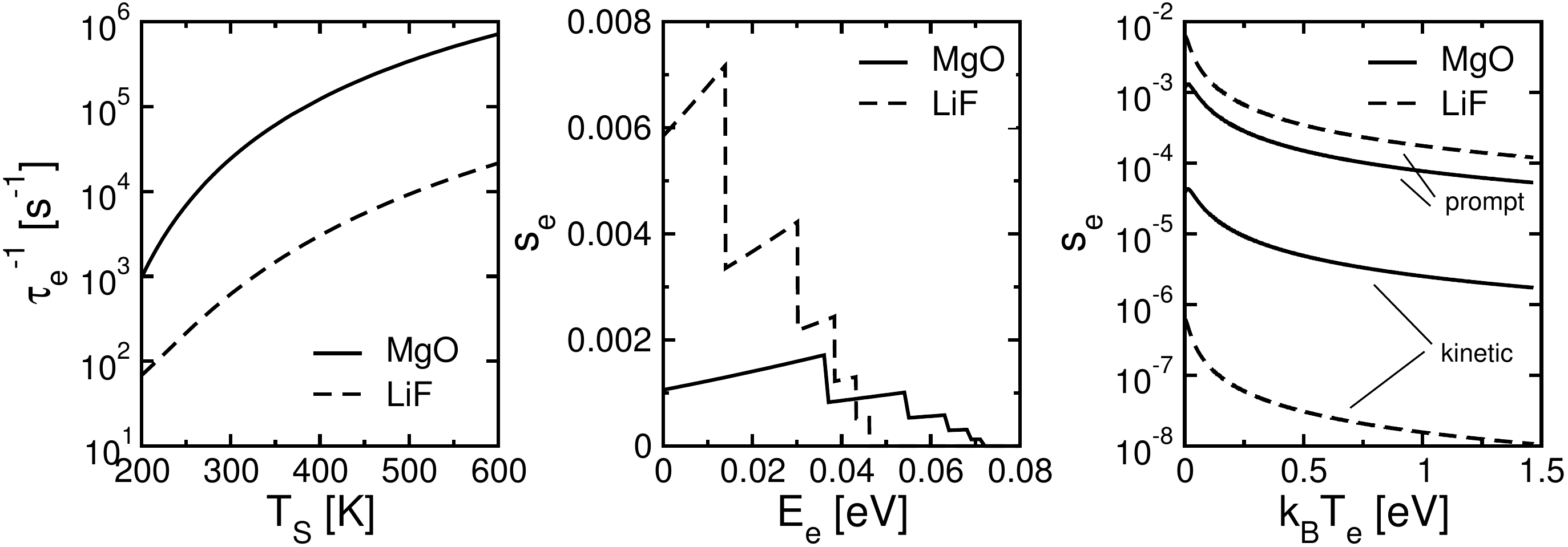}
\caption{Inverse desorption time $\tau_e^{-1}$ (left), prompt energy resolved sticking coefficient  (middle) and prompt as as well as kinetic energy averaged sticking coefficient (right) for MgO and LiF.}
\label{stick_des_overview}
\end{figure}

Electron physisorption at a dielectric surface with negative electron affinity is an intriguing phenomenon due to the interplay of potential depth, magnitude of matrix elements and surface temperature \cite{HBF10a,HBF10b,HBF11}.  Initial trapping of the electrons, characterised by the prompt sticking coefficient, occurs in the upper bound states by one-phonon transitions. Relaxation after initial trapping depends on the strength of transitions from the upper bound states to the lowest bound states. If the lowest bound state was linked to the second bound state by a one-phonon process the electron would relax for all surface temperatures. If they are linked by a multiphonon process relaxation takes place only for low temperature, whereas for room temperature a relaxation bottleneck ensues as the electron desorbs from the upper bound states before it drops to the lowest bound state. This leads to a severe reduction of the kinetic sticking coefficient compared to the prompt sticking coefficient.
 The 
dominant desorption channel depends also on the depth of the potential. For a shallow potential desorption 
occurs directly from the lowest bound state to the continuum. For deeper potentials desorption proceeds via 
the upper bound states. Desorption occurs then via a cascade in systems without and as a 
one-way process in systems with relaxation bottleneck~\cite{HBF11}. 
An overview of results for the desorption time as well as prompt and kinetic sticking coefficients for MgO and LiF is given by Fig. \ref{stick_des_overview}.
Most important for the plasma context is that \(s_e \ll 1\) and \(\tau_e^{-1}\neq 0\)~\cite{HBF10a,HBF10b,HBF11}, implying that a dielectric surface with a negative electron affinity is not a perfect absorber for electrons. 

\section{Mie scattering by a charged particle}
\label{mie}

In the previous section we have employed our model for the ESL to analyse the electron kinetics at the plasma boundary and to calculate desorption times and sticking coefficients for a dielectric with negative electron affinity. Another application of the model for the ESL is the 
study of the optical properties of the wall-bound electrons. In this section we will study the effect of surplus electrons on the scattering of light by a sphere (Mie scattering) with an eye on a charge measurement for dust particles.
 Most established methods for measuring the particle charge~\cite{CJG11,KRZ05,TLA00} require plasma parameters 
which are not precisely known. The particle charge, however, is the fundamental quantity of interest in the field of dusty plasmas. It determines the interactions between dust grains and between dust grains and external electromagnetic fields.
While Mie scattering is already used in the field of dusty plasmas to determine the particle size \cite{GCK12}, it has up to now not been used to determine the particle charge. 
Quite generally the Mie signal contains information about excess electrons as their
electrical conductivity modifies either the boundary condition for electromagnetic fields or the polarizability of
the material \cite{BH83,KK10,KK07,BH77}. But to what extent and in what spectral range the particle charge influences the scattering of light remains an open question.

The location of the surplus electrons depends on the electron affinity \(\chi\) of the dust grain.  For \(\chi<0\),  as it is the case for MgO, CaO or LiF \cite{RWK03,BKP07}, electrons are trapped in the image potential in front of the surface where they form a spherical two-dimensional electron gas around the grain. 
The image potential results from the static interaction of the electron to a surface mode associated with the transverse optical (TO) phonon. The residual dynamic interaction limits the surface conductivity \(\sigma_s\) of the image bound electrons. The electrical field parallel to the surface induces a surface current  \(\mathbf{K}=\sigma_s \mathbf{E}_\parallel\) which is proportional to the surface conductivity \(\sigma_s\). The surface current enters the 
boundary condition for the magnetic field. Thus, 
\(\hat{\mathbf{e}}_r \times \left( \mathbf{H}_i +\mathbf{H}_r -\mathbf{H}_t \right) =\frac{4\pi}{c}\mathbf{K}\), 
 where \(c\) is the speed of light \cite{BH77}.

 For \(\chi>0\), as it is the case for Al$_2$O$_3$ or SiO$_2$,  electrons  form a space charge layer in the conduction band \cite{HBF12}. Its width, limited by the screening in the dielectric, is typically larger than a micron. For micron sized particles we can thus assume a homogeneous distribution of the excess electrons in the bulk. The effect on light scattering is now encoded in the bulk conductivity of the excess electrons \(\sigma_b\), which is limited by scattering with a longitudinal optical (LO) bulk phonon.  The bulk conductivity gives rise to a polarizability \(\alpha=4\pi i \sigma_b /\omega\), with \(\omega\) the frequency, which enters the refractive index $N=\sqrt{\epsilon+\alpha}$.

To solve the scattering problem we expand the incident plane wave,
\begin{equation}
\mathbf{E}_i=E_0e^{ikz-i\omega t} \mathbf{\hat{e}}_x, \qquad \mathbf{H}_i=\frac{ck}{\omega} E_0 e^{ikz-i\omega t} \mathbf{\hat{e}}_y
\end{equation}
 in spherical vector harmonics $\mathbf{m}_{emn}$,$\mathbf{m}_{omn}$,  $\mathbf{n}_{emn}$, and $\mathbf{n}_{omn}$ \cite{Stratton41} which are solution of the vector wave equation $\nabla^2 \mathbf{C}+k^2 \mathbf{C}$=0 , in spherical coordinates
\begin{equation}
\mathbf{E}_i=E_0 e^{-i\omega t} \sum_{n=1}^\infty i^n \frac{2n+1}{n(n+1)} \left( \mathbf{m}_{o1n}^{(1)}-i\mathbf{n}_{e1n}^{(1)}\right),\end{equation}\begin{equation} \mathbf{H}_i=-\frac{c k }{\omega} E_0 e^{-i\omega t} \sum_{n=1}^\infty i^n \frac{2n+1}{n(n+1)} \left( \mathbf{m}_{e1n}^{(1)}+i\mathbf{n}_{o1n}^{(1)}\right)
\end{equation}
and match the incident wave with the reflected wave,
\begin{equation}
 \mathbf{E}_r=E_0 e^{-i\omega t}\sum_{n=1}^\infty i^n \frac{2n+1}{n(n+1)} \left(a_n^r \mathbf{m}_{o1n}^{(3)}-i b_n^r \mathbf{n}_{e1n}^{(3)} \right), 
\end{equation}
\begin{equation}
\mathbf{H}_r=-\frac{ck}{\omega} E_0 e^{-i\omega t} \sum_{n=1}^\infty i^n \frac{2n+1}{n(n+1)} \left( b_n^r \mathbf{m}_{e1n}^{(3)}+ia_n^r \mathbf{n}_{o1n}^{(3)}\right)
\end{equation}
and the transmitted wave,
\begin{equation}
 \mathbf{E}_t=E_0 e^{-i\omega t} \sum_{n=1}^\infty i^n \frac{2n+1}{n(n+1)} \left(a_n^t \mathbf{m}_{o1n}^{(1)}-i b_n^t \mathbf{n}_{e1n}^{(1)} \right),
\end{equation}
\begin{equation}
\mathbf{H}_t=-\frac{cq }{\omega }E_0 e^{-i\omega t} \sum_{n=1}^\infty i^n \frac{2n+1}{n(n+1)} \left( b_n^t \mathbf{m}_{e1n}^{(1)}+ia_n^t \mathbf{n}_{o1n}^{(1)}\right)
\end{equation}
(\(q\) is the wavenumber inside the particle) at the boundary of a dielectric sphere with radius \(a\). The boundary conditions  at the surface are given by
\begin{equation}
\mathbf{\hat{e}_r} \times \left( \mathbf{C}_i+\mathbf{C}_r-\mathbf{C}_t\right) =0
\end{equation} 
for \(\mathbf{C}=\mathbf{E,H}\) in the case of \(\chi>0\) and in the case of \(\chi<0\) by the above equation for \(\mathbf{E}\) and 
\begin{equation}
\hat{\mathbf{e}}_r \times \left( \mathbf{H}_i +\mathbf{H}_r -\mathbf{H}_t \right) =\frac{4\pi}{c}\sigma_s \mathbf{E}_\parallel . 
\end{equation}
This gives, following Bohren and Hunt \cite{BH77}, the scattering 
coefficients 
\begin{equation}
a_n^r=\frac{\psi_n(N\rho)\psi_n^\prime (\rho)-\left[N\psi_n^\prime(N\rho)-i\tau \psi_n(N\rho)\right] 
\psi_n(\rho) }{\left[N\psi_n^\prime(N\rho)-i\tau \psi_n(N\rho)\right]\xi_n(\rho)-\psi_n(N\rho)\xi_n^\prime(\rho)}
\end{equation}
and
\begin{equation}
b_n^r=\frac{\psi_n^\prime(N\rho)\psi_n(\rho)-\left[N\psi_n(N \rho)+i\tau \psi_n^\prime(N\rho)\right]\psi_n^\prime(\rho)}{\left[N\psi_n(N\rho)+i\tau \psi_n^\prime(N\rho) \right]\xi_n^\prime(\rho)-\psi_n^\prime(N\rho)\xi_n(\rho)}
\end{equation}
where for \(\chi<0\) (\(\chi>0\)) the dimensionless surface conductivity \(\tau(\omega)=4\pi \sigma_s(\omega)/c\) (\(\tau=0\)) and the refractive index \(N=\sqrt{\epsilon}\) (\(N=\sqrt{\epsilon+\alpha}\)), \(\rho= ka=2\pi a /\lambda\) the size parameter
, where \(k\) is the wavenumber, \(\psi_n(\rho)=\sqrt{\pi \rho/2} J_{n+1/2}(\rho)\),  
\(\xi_n(\rho)=\sqrt{\pi \rho/2} H^{(1)}_{n+1/2}(\rho)\) with \(J_n(\rho)\) the Bessel and \(H_n^{(1)}(\rho)\) 
the Hankel function of the first kind. 
As for uncharged particles the scattering efficiency is given by
\begin{equation}
Q_s=\frac{2}{\rho^2}\sum_{n=1}^\infty \left(2n+1\right)  \left(|a_n^r|^2+|b_n^r|^2 \right), 
\end{equation}
the extinction efficiency by
\begin{equation}
Q_t=-\frac{2}{\rho^2}\sum_{n=1}^\infty \left(2n+1\right) \Re \left(a_n^r+b_n^r \right)
\end{equation}
and the absorption efficiency by $Q_a=Q_t-Q_s$. Any effect of the surplus electrons on the scattering of light, encoded in \(a_n^r\) and \(b_n^r\), is due to the 
surface conductivity (\(\chi<0\)) or the bulk conductivity (\(\chi>0\)).

For $\chi<0$ we use a planar model, which we justify below, to calculate the surface conductivity.
For the calculation of the surface conductivity we will employ the $1/z$ image potential. The electron-phonon interaction is given by Eq. (\ref{EMHamiltonian}). We apply the unitary transformation $H\rightarrow UHU^{-1}$ with $U=e^{iS}$, $S=(i/\hbar\omega_s)\sum_{\mathbf{Q}_\parallel}a_{\mathbf{Q}_\parallel} \phi_{\mathbf{Q}_\parallel}-a_{\mathbf{Q}_\parallel}^\dagger \phi_{\mathbf{Q}_\parallel}^\ast $ to separate static and dynamic coupling \cite{Barton81}.  The former 
leads to the image potential \(V=-\Lambda_0 e^2 /z\) with \(\Lambda_0=(\epsilon_0-1)/(4(\epsilon_0+1))\) 
supporting a series of bound Rydberg states whose wave functions read 
\begin{equation}
\phi_{n \mathbf{k}}(\mathbf{x},z)=\frac{1}{\sqrt{A}}e^{i\mathbf{k}\mathbf{x}}\sqrt{\frac{\Lambda_0}{a_B n n!^2}} 
W_{n,1/2}\left(\frac{2\Lambda_0 z}{n a_B} \right) \label{wavefunction}
\end{equation}
with \(a_B\) the Bohr radius, \(\mathbf{k}=(k_x,k_y)\), \(\mathbf{x}=(x,y)\),  \(A\) the surface area and \(W_{n,m}\) Whittaker's function \cite{WW27}. 
 Since trapped electrons are thermalised with the surface and the spacing between
Rydberg states is large compared to $k_BT$, they occupy almost exclusively the lowest image band \(n=1\). Assuming 
a planar surface is justified provided the de Broglie wavelength \(\lambda_{dB}\) of the electron on the surface 
is smaller than the radius \(a\) of the sphere. For a surface electron with energy \(E_\mathrm{kin}/k_B=300 K \) one 
finds \(\lambda_{dB}\approx 8 \times 10^{-7} \) cm. Thus, for particle radii \(a>10 \) nm the plane-surface 
approximation is justified. The residual dynamic interaction enables momentum relaxation parallel to the surface 
and hence limits the surface conductivity. 
  Introducing 
annihilation (creation) operators \(c_\mathbf{k}^{(\dagger)}\) for electrons in the lowest image band, the Hamiltonian
describing the dynamic electron-phonon coupling in the lowest image band reads \cite{KI95}
\begin{equation}
H=\sum_\mathbf{k} \epsilon_\mathbf{k} c_\mathbf{k}^\dagger c_\mathbf{k} + \hbar \omega_s \sum_\mathbf{Q} 
a_\mathbf{Q}^\dagger a_\mathbf{Q}\nonumber +H_{int}
\end{equation}  with
\begin{equation}
H_{int}=\frac{1}{\sqrt{A}}\sum_{\mathbf{k},\mathbf{Q}}M_{\mathbf{k},\mathbf{Q}} c^\dagger_{\mathbf{k}+\mathbf{Q}}\left(a_\mathbf{Q}-a^\dagger_{-\mathbf{Q}} \right)c_\mathbf{k}
\end{equation}
where the matrix element, calculated with the wave function given by Eq. (\ref{wavefunction}), is 
\begin{equation}
M_{\mathbf{k} \mathbf{Q}}= \frac{2e\sqrt{\pi \Lambda_0 \hbar^3}}{m \sqrt{\omega_s Q}} 
\left( \frac{2 \Lambda_0}{Q a_B +2\Lambda_0} \right)^3  \left[\mathbf{Q}\cdot\mathbf{k}+\frac{Q^2}{2} \right] 
\end{equation}
($m$ is the electron mass).
Within the memory function approach \cite{GW72} the  surface conductivity can be written as 
\begin{equation}
\sigma_s(\omega)=\frac{e^2 n_s}{m} \frac{i}{\omega +M(\omega)} \label{surfcon}
\end{equation}
with \(n_s\) the surface electron density. The memory function is then evaluated up to second order in the electron phonon coupling \cite{HBF12b}. Since $M(\omega)$ is independent of $n_s$ the surface conductivity is proportional to the surface density of electrons $n_s$.

For \(\chi>0\) the interaction of the electron with a longitudinal optical bulk phonon limits the  bulk conductivity.  The coupling of the electron to this mode with frequency $\omega_{LO}$ is described by \cite{Mahan90}
\begin{equation}
H_{int}=\frac{1}{\sqrt{V}}\sum_{\mathbf{k},\mathbf{q}} \frac{M}{q} c_{\mathbf{k}+\mathbf{q}}^\dagger c_\mathbf{k} \left(a_\mathbf{q}+a_{-\mathbf{q}}^\dagger \right), 
\end{equation}
where \(M=\sqrt{2\pi e^2\hbar\omega_{LO}\left(\epsilon_\infty^{-1}-\epsilon_0^{-1} \right)}\).  For the calculation of the bulk conductivity we employ again the memory function approach. In this case the bulk conductivity is proportional to the electron density $n_b$.

\begin{figure}[t]
\includegraphics[width=\linewidth]{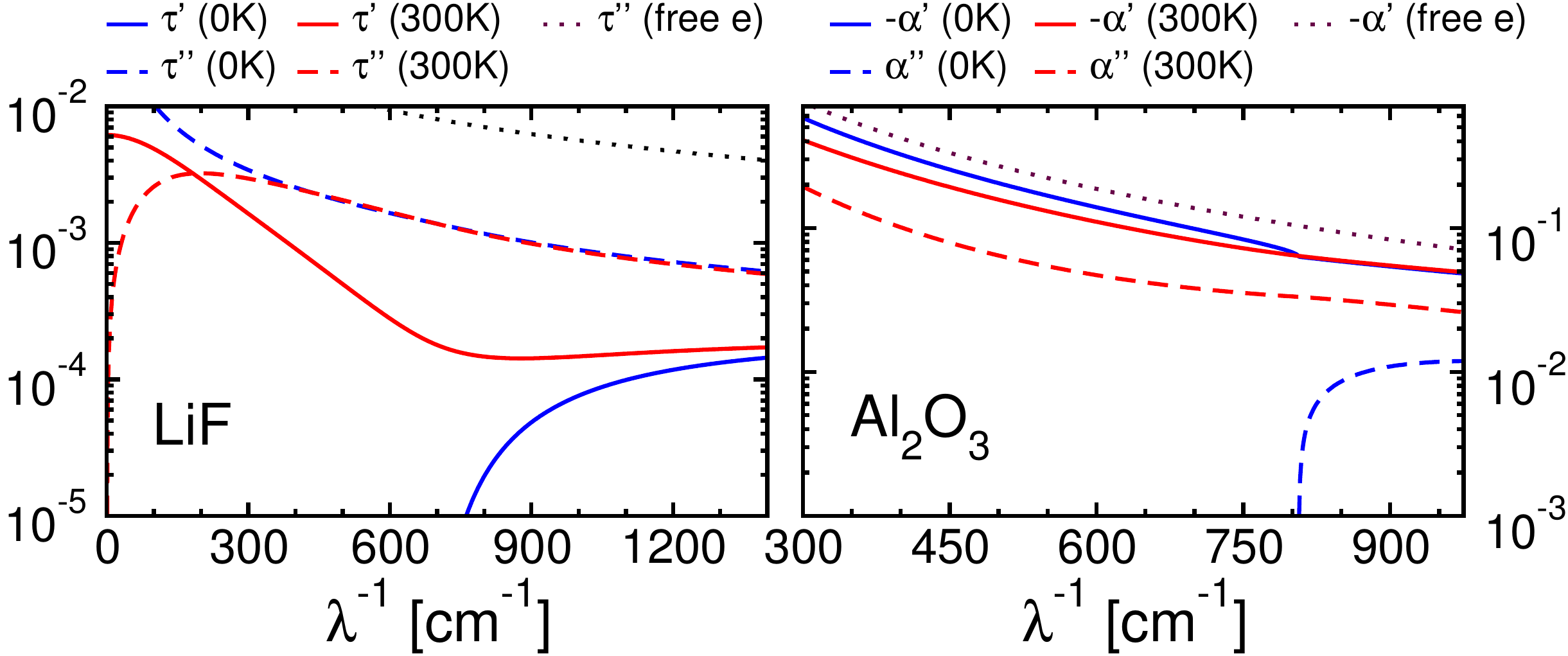}
\caption{Dimensionless surface conductivity \(\tau=\tau^{\prime}+i\tau^{\prime \prime}\) for LiF 
for \(n_s=10^{13}\) cm\(^{-2}\)  (left) and polarisability of excess electrons \(\alpha=\alpha^\prime+i\alpha^{\prime\prime}\) for Al$_2$O$_3$ 
for \(n_b=3\times 10^{17}\) cm\(^{-3}\) (right) as a function of the inverse wavelength \(\lambda^{-1}\).}
\label{cond_bis}
\end{figure}

The charge effect on scattering is encoded in the  surface conductivity for $\chi<0$ and the bulk conductivity for \(\chi>0\) which enter through the dimensionless surface conductivity $\tau$ or the polarisability $\alpha$ into the scattering coefficients. Figure \ref{cond_bis} shows $\tau=\tau^\prime+i\tau^{\prime \prime}$ for LiF and the polarisability for $\alpha=\alpha^\prime+\alpha^{\prime \prime}$ for Al$_2$O$_3$ as a function of the inverse wavelength $\lambda^{-1}$.
They turn out to be small even for a highly charged particle with \(n_s=10^{13}\) cm\(^{-2}\) (corresponding to \(n_b=3\times 10^{17}\) cm\(^{-3}\) for \(\chi>0\) and \(a=1\mu \)m). 
Compared to a free electron gas where \(M(\omega)=0\) (implying \(\tau^\prime=0\) and \(\alpha^{\prime \prime}=0\)), the electron-phonon coupling reduces \(\tau^{\prime \prime}\) and \(\alpha^\prime\) considerably.
 For \(T=0K\), \(\tau^\prime=0\) (\(\alpha^{\prime \prime}=0\)) for \(\lambda^{-1}<\lambda_s^{-1}=675\) cm\(^{-1}\), the inverse wavelength of the surface phonon ( \(\lambda^{-1}<\lambda_{LO}^{-1}=807\) cm\(^{-1}\), the inverse wavelength of the bulk LO phonon) since light absorption is only possible above the surface (bulk LO) phonon frequency. 
At room temperature \(\tau^{\prime \prime}\) and \(\alpha^\prime\) still outweigh \(\tau^\prime\) and \(\alpha^{\prime\prime}\). The temperature effect on  \(\tau^{\prime \prime}\) is less apparent for \(\lambda^{-1}>300\) cm\(^{-1}\) than for \(\alpha^\prime\) but for  \(\lambda^{-1}<300\) cm\(^{-1}\) a higher temperature lowers  \(\tau^{\prime \prime}\) considerably.

\begin{figure}[t]
%\sidecaption
{\includegraphics[width=\linewidth]{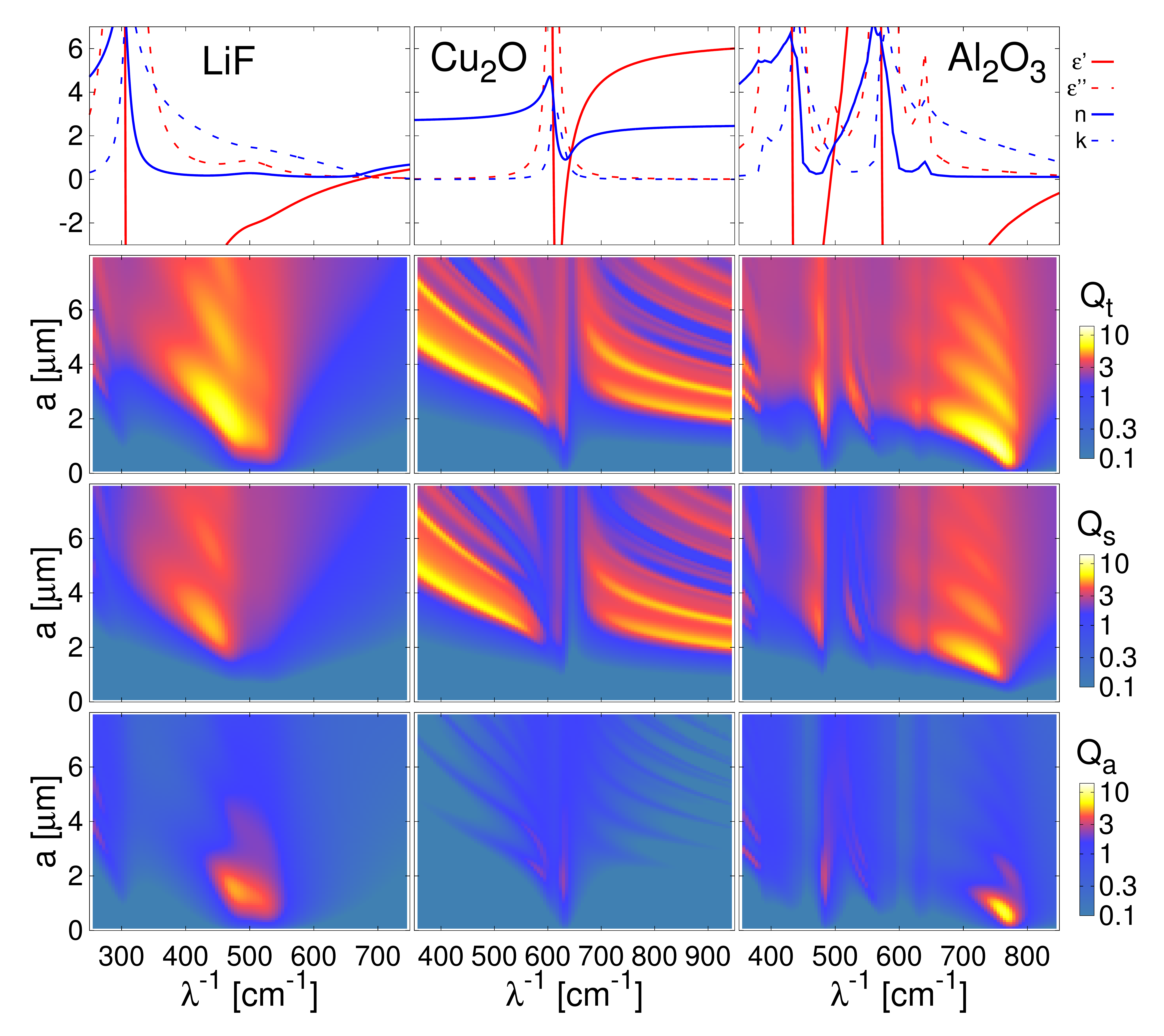}}
\caption{Dielectric constant \(\epsilon=\epsilon^\prime+ i \epsilon^{\prime \prime}\), 
refractive index \(N=n+ik\) (first row) and extinction efficiency \(Q_t\) (second row), scattering efficiency \(Q_s\) (third row) as well as absorption efficiency \(Q_a\) (fourth row) depending on the particle radius \(a\) for LiF, Cu$_2$O and Al$_2$O$_3$ as a function of 
\(\lambda^{-1}\).}
\label{mie2}
\end{figure}

We now turn to the scattering properties of the sphere. 
For LiF and Al$_2$O$_3$ one or two phonon modes dominate the dielectric constant. Far above the highest TO phonon mode (\(\lambda^{-1} > 700\) cm\(^{-1}\) for LiF and \(\lambda^{-1} > 900\) cm\(^{-1}\) for Al$_2$O$_3$) the refractive index $N$ is real, relatively small and does not vary strongly with frequency. In this regime a micron sized grain would give rise to a typical Mie plot exhibiting interference and ripples which are due to the complicated functional form of \(a_n^r\) and \(b_n^r\) and not due to the underlying dielectric constant.
Surplus electrons would moreover not alter the extinction behaviour in this region because \(|\epsilon| \gg |\tau|\) and \(|\epsilon|\gg|\alpha| \). 

However, immediately above the TO phonon resonance in $\epsilon$, $\epsilon^\prime <0$ and $\epsilon^{\prime \prime} \ll 1$. This allows for anomalous optical resonances which are sensitive to small variations of $\epsilon$ and also to $\tau$ and $\alpha$. They are due to resonant excitation of transverse surface modes of the sphere \cite{FK68} and have been first identified for metallic particles where they lie in the ultraviolet \cite{TL06,Tribelsky11}.
 For a dielectric the TO phonon induces them. 
Fig. \ref{mie2} shows the dielectric constant $\epsilon$, the refractive index $N$ as well as the extinction $Q_t$, scattering $Q_s$, and absorption efficiency $Q_a$ for LiF, Cu$_2$O and Al$_2$O$_3$ particles as a function of \(\lambda^{-1}\)  and the particle radius \(a\). For LiF and Al$_2$O$_3$ we find a clearly resolved series of optical resonances. For the smaller particles they are due to absorption while for larger particles they are scattering resonances. The crossover between scattering and absorption occurs in the first resonance. For comparison, we display in Fig. \ref{mie2} also results for Cu$_2$O particles which do not show clearly resolved resonances. For this material $\epsilon^\prime<0$ but $\epsilon^{\prime \prime}$ is not sufficiently small to allow for sharp resonances. Nevertheless for submicron-sized particles a small extinction resonance due to absorption can be identified near \(630\) cm\(^{-1}\).

\begin{figure}[t]
{\includegraphics[width=\linewidth]{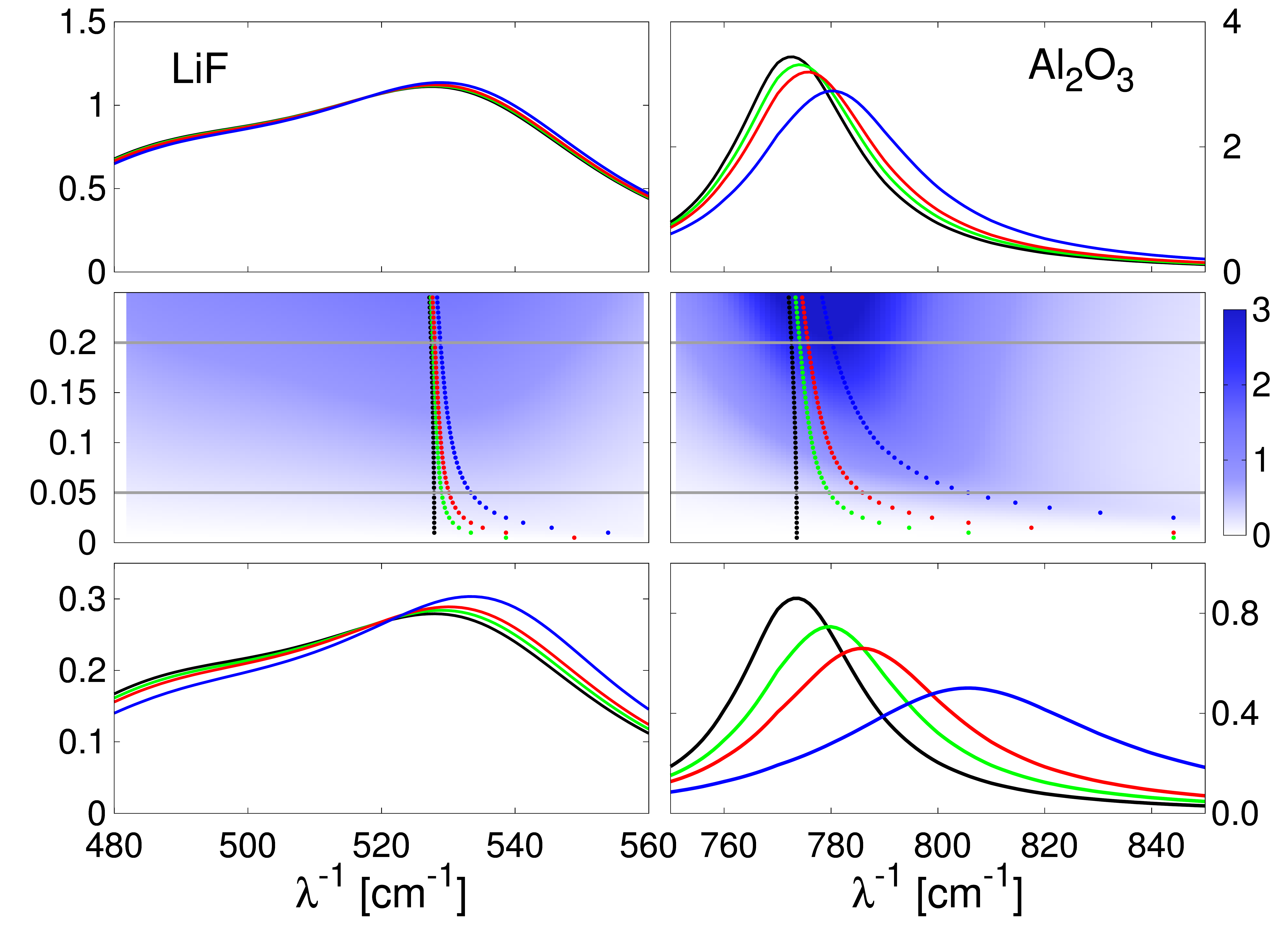}}
\caption{Middle panel: Extinction efficiency \(Q_t\) as a function of the inverse wavelength \(\lambda^{-1}\) and 
the radius \(a\) for a LiF with \(n_s= 5\times 10^{12} \)cm\(^{-2}\) (left)  and an Al$_2$O$_3$ particle with $n_b=3n_s/a$ (right) for \(T=300 K\). The 
dotted lines indicate the extinction maximum for \(n_s=0\) (black), 
\( 10^{12}\) (green), \(2\times 10^{12}\) (red), and \(5\times 10^{12}\) cm\(^{-2}\) (blue). Top and bottom panel: Extinction efficiency \(Q_t\) for different surface electron densities as a function of \(\lambda^{-1}\) and two radii \(a=0.2\mu\)m (top) and \(a=0.05\mu \)m (bottom).}
\label{mie3}
\end{figure}

The extinctions resonances are blue-shifted with increasing particle charge \cite{HBF12b}. This effect is most significant for small particles. Fig. \ref{mie3} shows the small particle tail of the lowest extinction resonance for LiF and Al$_2$O$_3$. The main panel shows the extinction efficiency as a function of the particle radius $a$ and the inverse wavelength $\lambda^{-1}$ for $n_s=5\times 10^{12}$cm$^{-1}$ (LiF) or $n_b=3n_s/a$ (Al$_2$O$_3$). Superimposed is the extinction maximum for several values of $n_s$ and corresponding $n_b$. The top and bottom panel show the lineshape of the extinction resonance for $a=0.2\mu$m and $a=0.05\mu$m. For both materials the surplus electrons lead to a blue-shift of the resonance. 

\begin{figure}[t]
\includegraphics[width=\linewidth]{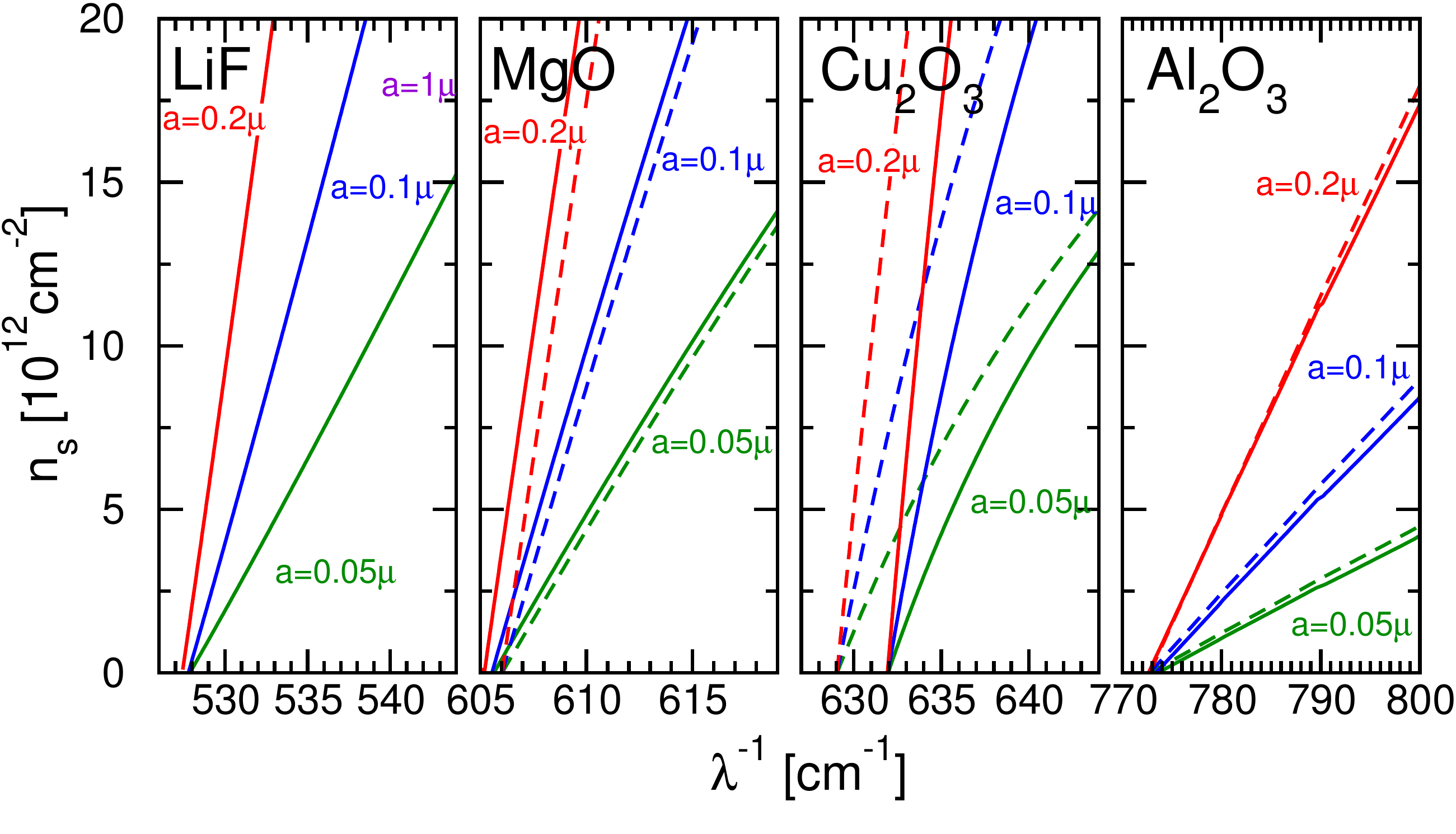}
\caption{Position of the extinction resonance depending on the surface charge $n_s$ for LiF, MgO, Cu$_2$O and Al$_2$O$_3$ (for equivalent bulk charge $n_b=3n_S/a$) particles with different radii $a$. Solid (dashed) lines are obtained from the Mie contour [Eqs. (\ref{resloca}) and (\ref{reslocb})].}
\label{miesum}
\end{figure}

For submicron-sized particles where the resonance shift is most significant $a<\lambda$. In this small particle limit we can expand the scattering coefficients for 
small \(\rho\). To ensure that in the limit of an uncharged surface, that is, for \(\tau \rightarrow 0\), \(a_n^r\) 
and \(b_n^r\) converge to their known small \(\rho\) expansions \cite{Stratton41}, we substitute \(t=\tau/\rho\) 
prior expanding the scattering coefficients. Up to \(\mathcal{O}(\rho^3)\) this yields \(a_1^r=a_2^r=b_2^r=0\) 
and only \(b_1^r\sim \mathcal{O}(\rho^3)\) contributes. Then the extinction efficiency reads
\begin{equation}
Q_t= \frac{12\rho\left(\epsilon^{\prime\prime}+\alpha^{\prime \prime}+2\tau^{\prime }/\rho \right)}
{\left(\epsilon^\prime+\alpha^\prime+2- 2\tau^{\prime\prime}/\rho \right)^2+
\left(\epsilon^{\prime \prime}+\alpha^{\prime \prime}+2 \tau^\prime/\rho \right)^2}~,
\end{equation}
where we have restored \(\tau\). Excess charges enter either by \(\tau\) (\(\chi<0\)) or \(\alpha\) (\(\chi>0\)). For \(\tau, \alpha \rightarrow 0\) this gives the limit of Rayleigh 
scattering. The resonance is located at  wavenumbers where
\begin{equation}
\epsilon^\prime+2- 2\tau^{\prime \prime}/\rho =0 \quad \mathrm{ for } \quad \chi<0 , \label{resloca} 
\end{equation}
\begin{equation}
\epsilon^\prime+\alpha^\prime +2 =0 \quad \mathrm{ for } \quad \chi>0 , \label{reslocb}
\end{equation}
and has a Lorentzian shape
 provided \(\epsilon^{\prime \prime}\) and \(\tau^\prime\) (or \(\alpha^{\prime\prime}\)) vary only 
 negligibly near the resonance wavelength. Figure \ref{mie3} confirms the Lorentzian lineshape for Al$_2$O$_3$. For LiF, however, \(\epsilon^{\prime \prime}\) has a hump (see Fig. \ref{mie2}) close to the resonance due to a second TO phonon mode which is much weaker than the dominant TO phonon. This leads to the deviation form the Lorentzian lineshape.

For an uncharged surface the resonance is at \(\lambda_0^{-1}\) for which \(\epsilon^\prime =-2\). For \(\chi<0\) the shift of the resonance is proportional to \(\tau^{\prime\prime}\) and thus to \(n_s\),
 provided \(\epsilon^\prime\) is well approximated linearly in \(\lambda^{-1}\) and \(\tau^{\prime \prime}\) does not vary 
significantly near \(\lambda_0^{-1}\). In this case, we substitute in (\ref{resloca}) the
expansions \(\epsilon^\prime=-2+c_\epsilon(\lambda^{-1}-\lambda_0^{-1})\) and \(\tau^{\prime \prime}=c_\tau n_s\) 
where \(c_\epsilon=\partial \epsilon^\prime/(\partial \lambda^{-1})|_{\lambda_0^{-1}}\) and 
\(c_\tau=\tau^{\prime \prime}/n_s|_{\lambda_0^{-1}}\). Then the resonance is located at 
\(\lambda^{-1}=\lambda_0^{-1}+c_\tau n_s/(\pi c_\epsilon a \lambda_0^{-1})\). For \(\chi>0\) the resonance is located at \(\lambda^{-1}=\lambda_0^{-1}-c_\alpha n_b/c_\epsilon \) where \(c_\alpha=\alpha^{\prime}/n_b|_{\lambda_0^{-1}}\). 

The proportionality of the resonance shift to \(n_s\) for LiF and \(n_b\) for Al$_2$O$_3$ can also be seen in
Fig. \ref{miesum} where we plot on the abscissa the shift of the extinction resonance arising 
from the surface electron density (or corresponding bulk electron density) given on the ordinate for LiF,  MgO (\(\chi<0\)) as well as Cu$_2$O and Al$_2$O$_3$ (\(\chi>0\)). Both bulk and surface electrons lead qualitatively to the same resonance shift. Note that even for Cu$_2$O  which does not show clearly resolved extinction resonance a shift is discernible. The most promising candidate for an optical charge measurement is Al$_2$O$_3$ where the shift is strongest.

 To illustrate the similarity of bulk and surface electron effects we consider the resonance condition (\ref{resloca}) or  (\ref{reslocb}) for free electrons, which then becomes
 \begin{equation}
 \epsilon^{\prime}-\frac{2Ne^2}{(m a^3\omega^2)}=-2
 \end{equation}
  for \(\chi<0\) and
  \begin{equation}
  \epsilon^\prime-\frac{3Ne^2}{(m^\ast a^3\omega^2)}=-2
  \end{equation} for \(\chi>0\), where \(N\) is the number of electrons on the sphere. The effect of surface electrons is weaker by a factor of \(2 m^\ast /3 m\). The factor \(2/3\) can be understood as a geometric factor as only the parallel component of the electric field acts on the spherically confined electron gas. Most important, however, the fact that \(\tau/\rho\) and \(\alpha\) enter the resonance condition on the same footing shows that the resonance blueshift is in the first place an electron density effect on the polarizability of the grain. We therefore expect the shift to prevail also for a more complex electron distribution on the grain between the two limiting cases of a surface and a homogeneous bulk charge.

\begin{figure}[t]
\includegraphics[width=\linewidth]{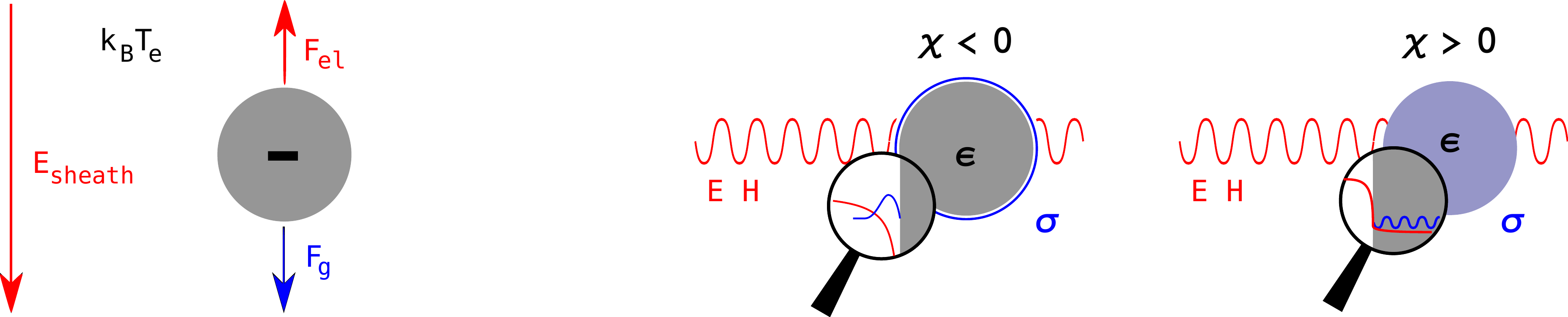}
\caption{Conventional charge measurement (left) and proposed optical charge measurement by Mie scattering (right). Conventional charge measurements rely on trapping the particle in the sheath and exploiting a force balance between gravity and the electric force on the particle. The knowledge of the plasma parameters is required to infer the particle charge. The optical measurement does not require the knowledge of plasma parameters. }
\label{mie_summary}
\end{figure}

Our results suggest to use the shift of the extinction resonance to determine the particles charge. This requires the use of particles with a strong TO phonon resonance in the dielectric constant \(\epsilon=\epsilon^\prime +i \epsilon^{\prime \prime}\)  which leads to \(\epsilon^\prime<0\) and \(\epsilon^{\prime \prime} \ll 1\) above the TO phonon frequency. The resoance shift is found for particles with surface  (\(\chi<0\), e.g. MgO, LiF) as well as bulk excess electrons (\(\chi>0\) e.g. Al$_2$O$_3$). For dusty plasmas an optical charge diagnostic can be rather attractive because established 
methods for measuring the particle charge~\cite{CJG11,KRZ05,TLA00} require plasma parameters 
which are not precisely known whereas the charge measurement by Mie scattering does not (see Fig. \ref{mie_summary}).
Particles showing the resonance shift could be employed as minimally invasive electric probes, which collect electrons depending on 
the local plasma environment. Determining their charge from Mie scattering and the forces acting on them 
by conventional means~\cite{CJG11,KRZ05,TLA00} would then allow to extract the local plasma parameters. Moreover, the Mie signal would provide a charge diagnostic for nanodust \cite{GCK12}, which is too small and light for traditional charge measurements.

\section{Summary}
\label{summ}

This chapter introduces a microscopic modelling of the plasma wall which complements traditional sheath models by an interface region---the ESL. In this model the negative charge on the plasma boundary is treated as a wall-thermalised electron distribution minimising the grand canonical potential and satisfying Poisson's equation. Its boundary with the plasma sheath is determined by a force balance between the attractive image potential and the repulsive sheath potential and lies in front of the crystallographic interface. Depending on the electron affinity \(\chi\), that is the offset of the conduction band minimum to the potential in front of the surface, two scenarios for the wall-bound electrons are realised. For \(\chi<0\) (e.g. MgO, LiF) electrons do not penetrate into the solid but are trapped in the image states in front of the surface where they form a quasi two-dimensional electron gas. For \(\chi>0\) (e.g. SiO$_2$, Al$_2$O$_3$) electrons penetrate into the conduction band where they form an extended space charge.

These different scenarios are also reflected in the electron physisorption at the wall. For \(\chi<0\) electrons from the plasma cannot penetrate into the solid. They are trapped in the image states in front of the surface. The transitions between unbound and bound states are due to surface vibrations. In this case the sticking coefficient for electrons is relatively small, typically on the order of $10^{-3}$. 
For $\chi>0$ electron physisorption takes place in the conduction band. For this case sticking coefficients and desorption times have not been calculated yet but in view of the more efficient scattering with bulk phonons, responsible for electron energy relaxation in this case, we expect them to be larger than for the case of $\chi<0$.
This indicates that the electron affinity should be an important parameter  for the charging-up of plasma walls. In particular dusty plasmas offer the possibility to study experimentally the charge-up of dust grains depending on this parameter.

Our microscopic model can also be applied to study the optical properties of the wall-bound electrons on a dust particle. Surplus electrons affect the polarisability of the dust particle by their surface (\(\chi <0\)) or bulk conductivity (\(\chi>0\)). This leads to a blue-shift of an extinction resonance in the infrared for negatively charged dust particles. This effect offers an optical way to measure the particle charge which, unlike tradition charge measurements, does not require the plasma parameters at the position of the particle.

\begin{acknowledgement}
This work was supported by Deutsche Forschungsgemeinschaft through SFB-TR 24.
\end{acknowledgement}


\begin{thebibliography}{10}
\providecommand{\url}[1]{{#1}}
\providecommand{\urlprefix}{URL }
\expandafter\ifx\csname urlstyle\endcsname\relax
  \providecommand{\doi}[1]{DOI \discretionary{}{}{}#1}\else
  \providecommand{\doi}{DOI \discretionary{}{}{}\begingroup
  \urlstyle{rm}\Url}\fi

\bibitem{Franklin76}
R.N. Franklin, \emph{Plasma phenomena in gas discharges} (Clarendon Press,
  Oxford, 1976)

\bibitem{WTE10}
C.J. Wagner, P.A. Tschertian, J.G. Eden, Appl. Phys. Lett. \textbf{97}, 134102
  (2010)

\bibitem{BBB09}
H.~Baumgartner, D.~Block, M.~Bonitz, Contrib. Plasma Phys. \textbf{49}, 281
  (2009)

\bibitem{PAB08}
A.~Piel, O.~Arp, D.~Block, I.~Pilch, T.~Trottenberg, S.~Kaeding, A.~Melzer,
  H.~Baumgartner, C.~Henning, M.~Bonitz, Plasma Phys. Control. Fusion
  \textbf{49}, 281 (2008)

\bibitem{FIK05}
V.E. Fortov, A.V. Ivlev, S.A. Khrapak, A.G. Khrapak, G.E. Morfill, Phys. Rep.
  \textbf{421}, 1 (2005)

\bibitem{BKS06}
J.~Berndt, E.~Kova\v{c}evi\'{c}, V.~Selenin, I.~Stefanovi\'{c}, J.~Winter,
  Plasma Sources Sci. Technol. \textbf{15}, 18 (2006)

\bibitem{GMB02}
Y.B. Golubovskii, V.A. Maiorov, J.~Behnke, J.F. Behnke, J. Phys. D: Appl. Phys
  \textbf{35}, 751 (2002)

\bibitem{WYB05}
H.E. Wagner, Y.V. Yurgelenas, R.~Brandenburg, Plasma Phys. Control. Fusion
  \textbf{47}, B641 (2005)

\bibitem{BMG05}
R.~Brandenburg, V.A. Maiorov, Y.B. Golubovskii, H.E. Wagner, J.~Behnke, J.F.
  Behnke, J. Phys. D: Appl. Phys \textbf{38}, 2187 (2005)

\bibitem{SLP07}
L.~Stollenwerk, J.G. Laven, H.G. Purwins, Phys. Rev. Lett. \textbf{98}, 255001
  (2007)

\bibitem{BWS12}
M.~Bogaczyk, R.~Wild, L.~Stollenwerk, H.E. Wagner, J. Phys. D: Appl. Phys
  \textbf{46}, 465202 (2012)

\bibitem{BNW12}
M.~Bogaczyk, S.~Nemschokmichal, R.~Wild, L.~Stollenwerk, R.~Brandenburg,
  J.~Meichsner, H.E. Wagner, Contrib. Plasma Phys. \textbf{52}, 847 (2012)

\bibitem{BSE06}
K.H. Becker, K.H. Schoenbach, J.G. Eden, J. Phys. D: Appl. Phys \textbf{39},
  R55 (2006)

\bibitem{Kushner05}
M.J. Kushner, J. Phys. D: Appl. Phys \textbf{38}, 1633 (2005)

\bibitem{Kogelschatz03}
U.~Kogelschatz, Plasma Chemistry and Plasma Processing \textbf{23}, 1 (2003)

\bibitem{THW11}
P.A. Tschertian, C.J. Wagner, T.L. Houlahan, B.~Li, D.J. Sievers, J.G. Eden,
  Contrib. Plasma Phys. \textbf{51}, 889 (2011)

\bibitem{DOL10}
R.~Dussart, L.~Overzet, P.~Lefaucheux, T.~Dufour, M.~Kulsreshath, M.~Mandra,
  T.~Tillocher, O.~Aubry, S.~Dozias, P.~Ranson, J.~Lee, , M.~Goeckner, Eur.
  Phys. J. D \textbf{60}, 601 (2010)

\bibitem{EC87}
K.G. Emeleus, J.R.M. Coulter, Int. J. Electronics \textbf{62}, 225 (1987)

\bibitem{BBD97}
J.F. Behnke, T.~Bindemann, H.~Deutsch, K.~Becker, Contrib. Plasma Phys.
  \textbf{37}, 345 (1997)

\bibitem{HBF12}
R.L. Heinisch, F.X. Bronold, H.~Fehske, Phys. Rev. B \textbf{85}, 075323 (2012)

\bibitem{Stern78}
F.~Stern, Phys. Rev. B \textbf{17}, 5009 (1978)

\bibitem{TD85}
E.E. Tkharev, A.L. Danilyuk, Vacuum \textbf{35}, 183 (1985)

\bibitem{LL05}
M.A. Lieberman, A.J. Lichtenberg, \emph{Principles of plasma discharges and
  materials processing} (Wiley-Interscience, New York, 2005)

\bibitem{Riemann91}
K.U. Riemann, J. Phys. D: Appl. Phys \textbf{24}, 493 (1991)

\bibitem{Jackson98}
J.D. Jackson, \emph{Classical electrodynamics} (Wiley, New York, 1998)

\bibitem{CRL98}
J.B. Cui, J.~Ristein, L.~Ley, Phys. Rev. Lett. \textbf{81}, 429 (1998)

\bibitem{MRL01}
F.~Maier, J.~Ristein, L.~Ley, Phys. Rev. B \textbf{64}, 165411 (2001)

\bibitem{SS84}
F.~Stern, S.D. Sarma, Phys. Rev. B \textbf{30}, 840 (1984)

\bibitem{Lueth}
H.~L\"{u}th, \emph{Solid Surfaces, Interfaces and Thin Films} (Springer-Verlag,
  1992)

\bibitem{KS65}
W.~Kohn, L.J. Sham, Phys. Rev. \textbf{140A}, 1133 (1965)

\bibitem{Mermin65}
N.D. Mermin, Phys. Rev. \textbf{137A}, 1441 (1965)

\bibitem{EC88}
K.G. Emeleus, J.R.M. Coulter, IEE Proceedings \textbf{135}, 76 (1988)

\bibitem{BFHM12}
F.X. Bronold, H.~Fehske, R.L. Heinisch, J.~Marbach, Contrib. Plasma Phys.
  \textbf{52}, 859 (2012)

\bibitem{HBF10a}
R.L. Heinisch, F.X. Bronold, H.~Fehske, Phys. Rev. B \textbf{81}, 155420 (2010)

\bibitem{HBF10b}
R.L. Heinisch, F.X. Bronold, H.~Fehske, Phys. Rev. B \textbf{82}, 125408 (2010)

\bibitem{HBF11}
R.L. Heinisch, F.X. Bronold, H.~Fehske, Phys. Rev. B \textbf{83}, 195407 (2011)

\bibitem{GKT80a}
Z.W. Gortel, H.J. Kreuzer, R.~Teshima, Phys. Rev. B \textbf{22}, 5655 (1980)

\bibitem{KG86}
H.J. Kreuzer, Z.W. Gortel, \emph{Physisorption Kinetics} (Springer Verlag,
  Berlin, 1986)

\bibitem{IN76}
G.~Iche, P.~Nozi\'{e}res, J. Phys. (Paris) \textbf{37}, 1313 (1976)

\bibitem{Brenig82}
W.~Brenig, Z. Phys. B \textbf{48}, 127 (1982)

\bibitem{EM73}
E.~Evans, D.L. Mills, Phys. Rev. B \textbf{8}, 4004 (1973)

\bibitem{Barton81}
G.~Barton, J. Phys. C \textbf{14}, 3975 (1981)

\bibitem{BY73}
B.~Bendow, S.C. Ying, Phys. Rev. B \textbf{7}, 622 (1973)

\bibitem{CJG11}
J.~Carstensen, H.~Jung, F.~Greiner, A.~Piel, Phys. Plasmas \textbf{18}, 033701
  (2011)

\bibitem{KRZ05}
S.A. Khrapak, S.V. Ratynskaia, A.V. Zobnin, A.D. Usachev, V.V. Yaroshenko, M.H.
  Thoma, M.~Kretschmer, H.~Hoefner, G.E. Morfill, O.F. Petrov, V.E. Fortov,
  Phys. Rev. E \textbf{72}, 016406 (2005)

\bibitem{TLA00}
E.B. Tomme, D.A. Law, B.M. Annaratone, J.E. Allen, Phys. Rev. Lett.
  \textbf{85}, 2518 (2000)

\bibitem{GCK12}
F.~Greiner, J.~Carstensen, N.~Koehler, I.~Pilch, H.~Ketelsen, S.~Knist,
  A.~Piel, Plasma Sources Sci. Technol. \textbf{21}, 065005 (2012)

\bibitem{BH83}
C.F. Bohren, D.R. Huffman, \emph{Absorption and Scattering of Light by small
  particles} (Wiley, 1983)

\bibitem{KK10}
J.~Kla\v{c}ka, M.~Kocifaj, Progress in Electromagnetics Research \textbf{109},
  17 (2010)

\bibitem{KK07}
J.~Kla\v{c}ka, M.~Kocifaj, J. of Quantitative Spectroscopy and Radiative
  Transfer \textbf{106}, 170 (2007)

\bibitem{BH77}
C.F. Bohren, A.J. Hunt, Can. J. Phys. \textbf{55}, 1930 (1977)

\bibitem{RWK03}
M.~Rohlfing, N.P. Wang, P.~Krueger, J.~Pollmann, Phys. Rev. Lett. \textbf{91},
  256802 (2003)

\bibitem{BKP07}
B.~Baumeier, P.~Krueger, J.~Pollmann, Phys. Rev. B \textbf{76}, 205404 (2007)

\bibitem{Stratton41}
J.A. Stratton, \emph{Electromagnetic theory} (McGraw-Hill, 1941)

\bibitem{WW27}
E.T. Whittaker, G.N. Watson, \emph{A course of modern analysis} (Cambridge
  University Press, 1927)

\bibitem{KI95}
M.~Kato, A.~Ishii, Appl. Surf. Sci. \textbf{85}, 69 (1995)

\bibitem{GW72}
W.~G\"otze, P.~W\"{o}lfle, Phys. Rev. B \textbf{6}, 1226 (1972)

\bibitem{HBF12b}
R.L. Heinisch, F.X. Bronold, H.~Fehske, Phys. Rev. Lett. \textbf{109}, 243903
  (2012)

\bibitem{Mahan90}
G.D. Mahan, \emph{Many-particle physics} (Plenum, 1990).
\newblock P. 703-708

\bibitem{FK68}
R.~Fuchs, K.L. Kliewer, J. Opt. Soc. Am. \textbf{58}, 319 (1968)

\bibitem{TL06}
M.I. Tribelsky, B.S. Luk'yanchuk, Phys. Rev. Lett. \textbf{97}, 263902 (2006)

\bibitem{Tribelsky11}
M.I. Tribelsky, Europhys. Lett. \textbf{94}, 14004 (2011)

\end{thebibliography}
\end{document}